\documentclass[useAMS,usenatbib]{mn2e}

\newcommand{\beq}{\begin{equation}}              
\newcommand{\eeq}{\end{equation}}                
\newcommand{\LCDM}{\Lambda{\rm CDM}}
\newcommand{\kms}{{{\rm \,km~s^{-1}}}}
\newcommand{\Mvir}{M_{\rm vir}}
\newcommand{\rvir}{r_{\rm vir}}
\newcommand{\Mstars}{M_{\star}}
\newcommand{\Msun}{{\rm M}_{\odot}}

\newcommand{\vopt}{v_{\rm opt}}
\newcommand{\vvir}{v_{\rm vir}}
\newcommand{\cvir}{c_{\rm vir}}

\title
[Coevolution of galaxies and dark haloes]
{The coevolution of the velocity and mass functions of galaxies and dark haloes}

\author[K.-H. Chae]
{Kyu-Hyun~Chae$^{1,2\star}$\\
\\
$^1$Sejong University, Department of Astronomy and Space Science, 
 98 Gunja-dong, Gwangjin-Gu, Seoul 143-747, Republic of Korea\\
$^2$Theoretical Astrophysics, Fermi National Accelerator Laboratory, P.~O. Box 
500, Batavia, IL 60510, USA\\
$^\star$chae@sejong.ac.kr: On sabbatical leave at Fermilab}
\date{
Accepted ........
Received .......;
in original form ......}

\pubyear{2010}

\begin{document}
\maketitle

\begin{abstract}
We employ a  bias-corrected abundance matching technique to investigate 
the coevolution of the $\LCDM$ dark halo mass function (HMF), the 
observationally derived velocity dispersion and stellar mass functions 
(VDF, SMF) of galaxies between $z=1$ and $0$. 
We use for the first time the evolution of the VDF constrained through strong 
lensing statistics by Chae (2010) for galaxy-halo abundance matching studies. 
As a local benchmark we use a couple of $z \sim 0$ VDFs (a Monte-Carlo 
realised VDF based on SDSS DR5 and a directly measured VDF based on SDSS DR6). 
We then focus on connecting the VDF evolution to the HMF evolution predicted 
by $N$-body simulations and the SMF evolution constrained by galaxy surveys.
On the VDF-HMF connection, 
we find that the local dark halo virial mass-central stellar 
velocity dispersion ($\Mvir$-$\sigma$) relation is in good agreement with
the individual properties of well-studied low-redshift dark haloes, and 
the VDF evolution closely parallels the HMF evolution meaning little evolution 
in the $\Mvir$-$\sigma$ relation. On the VDF-SMF connection, it is also likely 
that the stellar mass-stellar velocity dispersion ($\Mstars$-$\sigma$) relation 
 evolves little taking the abundance matching results together with
 other independent observational results and hydrodynamic simulation results. 
 Our results support the simple picture that as the halo grows hierarchically,
 the stellar mass and the central stellar velocity dispersion grow in parallel. 
We discuss possible implications of this parallel coevolution for galaxy
formation and evolution under the $\LCDM$ paradigm.

\end{abstract}
\begin{keywords}  galaxies: evolution -- galaxies: formation -- galaxies: haloes
-- galaxies: kinematics and dynamics -- galaxies: statistics -- 
galaxies: structure
\end{keywords}

\maketitle

\section{Introduction}

The current $\LCDM$ hierarchical structure formation theory predicts robustly 
the evolution of the dark halo mass\footnote{Throughout this refers to the 
total mass within the virial radius of the halo. Accordingly, it includes the 
stellar mass once the galaxy is formed.} function (HMF) over cosmic time 
(e.g., \citealt{Spr05,War06,Ree07,Luk07,Tin08,Kly10}). Because visible galaxies
are believed to form and reside within the haloes under the $\LCDM$ paradigm, 
the statistical functions of galaxies, such as the luminosity function (LF), 
the stellar mass function (SMF), and the stellar velocity  (dispersion) 
function (VF, VDF), are also expected to evolve. 
Connection of these statistical functions of galaxies with the theoretical HMF
is not straightforward due to, and mirrors, the complex processes of galaxy 
formation and evolution including star formations, supernovae explosions, AGN 
activities and galaxy merging. Some recent works in the literature are focused 
on the connection of the HMF with the broadly measured stellar mass function 
(SMF) of galaxies (e.g., \citealt{CW09,Mos09,Guo09}). 
 Notice that the SMF as well as the LF have mainly to do with the star 
formation history of galaxies.

Galaxy formation in the halo has dynamical consequences as well. As stars are
formed in the inner halo, the halo responds and the inner
halo dark matter distribution is modified 
(e.g., \citealt{Blu86,Gne04,Rud08,Aba09,Tis10}).
Consequently, not only the total (i.e.\ dark plus stellar) mass distribution
but also the dark matter distribution may become different from the pure
dark matter distribution predicted by the $\LCDM$. This dynamical aspect of
galaxy formation is a crucial part of cosmological studies. Ultimately, the 
theory of galaxy formation should predict successfully the dynamical evolution
as well as the star formation history of galaxies. The statistical property of
the dynamics of galaxies is encoded in the VDF of galaxies. The local total 
VDF is carefully reconstructed by \citet{Cha10} using SDSS DR5 galaxy counts 
and intrinsic correlations between luminosities and velocities of galaxies.
More recently, \citet{Ber10} estimates directly the local total VDF based on 
SDSS DR6 measurements (of a DR4 sample).  \citet{Cha10} 
then constrains the evolution of the VDF up to $z \sim 1$ through the 
statistical properties of strong lensing galaxies based on the empirical result
that the average total (luminous plus dark) mass profile of galaxies is 
isothermal in the optical region. 
\citet{Cha10} notices that the {\it differential} evolution of the 
derived VDF is qualitatively similar to the evolution of the theoretical 
HMF under the current $\LCDM$ paradigm.

In this work we make a detailed quantitative comparison between the VDF 
evolution constrained from strong lensing statistics through the method of
\citet{Cha10} and the evolutions of mass functions (i.e.\ the HMF and the SMF) 
from the literature. 
In doing so, we investigate the local (statistical) 
correlations of the velocity dispersion ($\sigma$) of a galaxy with the virial 
mass ($\Mvir$) of the surrounding halo and the stellar mass ($\Mstars$) of the 
galaxy, i.e.\ $\sigma(\Mvir)$ and $\sigma(\Mstars)$, and their evolutions
out to $z \sim 1$. These empirical correlations will provide independent 
{\it statistical} constraints on the structures of galaxies and haloes and 
their evolutions. We investigate the implications of the local correlations
for the baryon-modified dark halo structures in a following work. In this work
we focus on the evolutions of the correlations. 
 We find that the halo mass, the stellar mass and the stellar velocity 
dispersion are coevolving in a parallel way for $0 \la z \la 1$. 
We discuss the implications of this result for galaxy 
formation and evolution under the $\LCDM$ paradigm.

This paper is organised as follows. In §2, we describe the method of analysis 
and the statistical functions to be used in this work; 
 some details of the analysis are given in the Appendix~A.
In \S 3, we investigate the connection between the (evolving) VDF of galaxies 
from the SDSS and strong lensing statistics and the HMF from N-body simulations.
 We obtain the relation $\sigma(\Mvir)$ and its evolution. 
We also examine the compatibility of the evolutions of the VDF and the HMF. 
In \S 4, we compare the VDF evolution with the SMF evolution  from galaxy 
surveys. We investigate the evolution of $\sigma(\Mstars)$ and the 
compatibility of the current observationally constrained VDF and SMF. 
In \S 5, we discuss the implications of the results for galaxy formation 
and evolution in the context of the current $\LCDM$ structure formation 
paradigm and cosmological observations. We conclude in \S 6.
Unless specified otherwise, we assume a WMAP 5 year $\LCDM$ 
cosmology (\citealt{Dun09}) with 
$(\Omega_{{\rm m}0},\Omega_{{\Lambda}0})=(0.25,0.75)$ and $H_0 =100h\kms$~Mpc$^{-1}$.
When parameter $h$ does not appear explicitly, $h=0.7$ is assumed.
In Appendix~B we compare the results of this work with the nearly 
concurrent results by \citet{Dut10}. \citet{Dut10} focus on the 
connection between the circular velocity in the optical region 
(at about the projected half-light radius) $\vopt$ and that at 
the virial radius $\vvir$. While \citet{Dut10} use various estimates of the
halo mass including satellite kinematics, weak lensing and abundance matching, 
their results are confined to $z \sim 0$. Another important difference 
between \citet{Dut10} and this work is that \citet{Dut10} use observed stellar 
mass-velocity relations while this work uses  the observationally
derived velocity dispersion functions for abundance matching.

\section{Method: abundance matching of statistical functions}

A statistical function of galaxies or haloes at a given epoch is defined by
\beq
\phi (x) = \left| \frac{dn(>x)}{dx} \right|,
\label{Func}
\eeq
where $x$ is the variable under consideration (e.g.\ $\sigma$, $\Mstars$, 
$\Mvir$) and $n(>x)$ is the integrated comoving number density down to $x$.

We use the abundance matching method (e.g.\ \citealt{Kra04,VO04,CWK06}) to 
relate statistically one variable ($x$) to another ($y$). Namely, we have
\beq
y=y(x) \hspace{1em} {\mbox{or}} \hspace{1em}  x=x(y) \hspace{1em} 
{\mbox{from}} \hspace{1em} n(>x)=n(>y).
\label{AM}
\eeq 
The key assumption for equation~(\ref{AM}) to be valid is that the two variables
 are monotonically increasing functions of each other.
 The accuracy of the median 
relation between $x$ and $y$ derived from the abundance matching method depends
on the nature of the intrinsic scatter of the true relation 
(see \citealt{Tas04,Beh10}). In Appendix~A, 
a simulation is carried out to investigate the possible effect of the intrinsic
scatter. It turns out that based on an observationally motivated intrinsic 
scatter  the abundance matching method reproduces the overall
behaviour of the intrinsic relation up to a maximum bias of $\sim 0.08$~dex
in most cases. We estimate and correct the biases in our abundance matching
analyses.

\begin{figure}
\begin{center}
\setlength{\unitlength}{1cm}
\begin{picture}(9,11)(0,0)
\put(-0.3,-0.4){\includegraphics{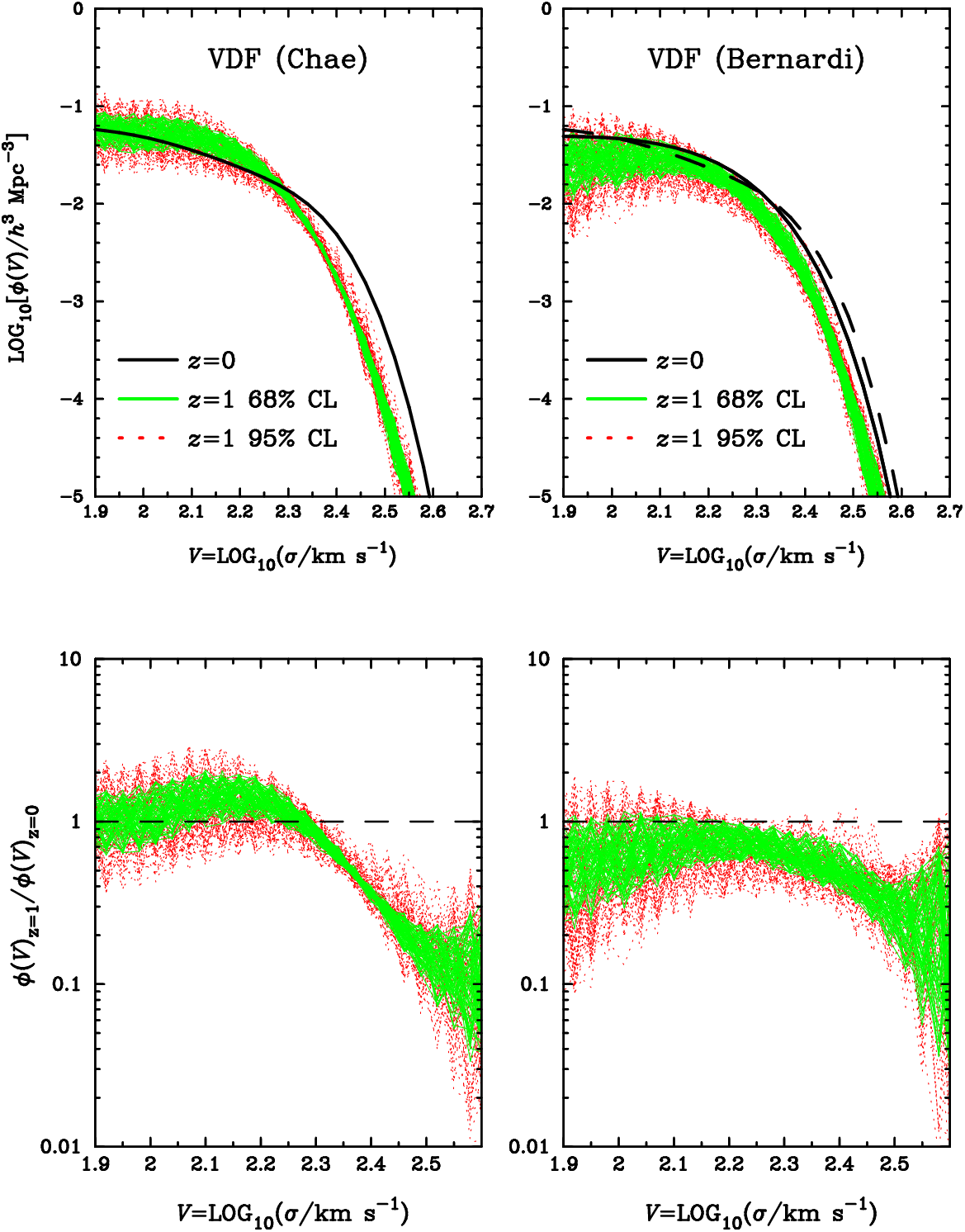}}
\end{picture}
\caption{Velocity dispersion functions (VDFs) of galaxies at $z=0$ and $z=1$. 
The VDFs at $z=0$ are those from \citet{Cha10} and \citet{Ber10}. The VDF by 
\citet{Cha10} is a sum of the early-type and late-type VDFs based on SDSS DR5 
data. The VDF by \citet{Ber10} is a direct fit for all galaxies based on SDSS 
DR6. The VDF by \citet{Cha10} is also reproduced as a dashed curve on the 
upper right panel for comparison.
The constraints on the VDFs at $z=1$ are based on strong lensing statistics
described in \citet{Cha10}.   Notice that strong lensing data
probe only the range of $95\kms \la \sigma \la 300 \kms$ 
(see the texts in \S~2).
}
\label{VDFs}
\end{center}
\end{figure}

This work is primarily concerned with connecting the stellar velocity dispersion
 of a galaxy ($\sigma$) to the dark halo virial  mass ($\Mvir$) and the galaxy 
stellar mass ($\Mstars$). For the VDF at $z=0$ we use the results from 
\citet{Cha10} and \citet{Ber10} (Fig.~\ref{VDFs}). 
Specifically, we use the `A0' VDF of \citet{Cha10} that is 
a result from combining the early-type and the late-type VDFs based on SDSS DR5
data. The \citet{Ber10} VDF is a {\it direct} fit to all-type galaxies based on 
 SDSS DR6.  \citet{Ber10} give various 
fit results depending on the range of velocity dispersion and fit method. 
For this work we use the fit result for $\sigma > 125 \kms$ (taking 
into account measurement errors) because this work is primarily concerned with
the evolution of massive galaxies. We then use strong lensing 
statistical analysis of \citet{Cha10} to constrain the evolution of the VDF 
up to $z=1$. The results are shown in Fig.~\ref{VDFs}.  
The result for the \citet{Cha10} VDF is a reproduction  
while that for the \citet{Ber10} VDF is a new result. Notice that for the 
\citet{Ber10} VDF a modified Schechter function introduced by \citet{She03} is 
used while for the \citet{Cha10} VDF a correction term is included.
The simplicity of the \citet{Ber10} VDF allows all four parameters of the 
function to be varied and constrained by strong lensing data in contrast to the
\citet{Cha10} VDF for which some parameters must be fixed (see \S 5.3 of 
\citealt{Cha10}). 
 
Notice that the strong lensing surveys used to constrain the evolution of the 
VDF are limited to lensing galaxies with image splitting greater than 0.3~arcsec
in the redshift range of $0.3 \la z \la 1$ (see \citealt{Cha10}). This lower
limit on image splitting implies that the constrained evolution of the VDF is
strictly valid only for $\sigma \ga 95\kms$. This in turn corresponds to
 $\Mstars \ga 10^{10.2}\Msun$ and $\Mvir \ga 10^{11.6} \Msun$ as will be shown
below. Furthermore, although the surveys do not have physically meaningful
upper limits on image separations, the surveys (because of the small sample
sizes) have only identified lensing galaxies with measured or implied stellar 
velocity dispersion $\sigma \la 300 \kms$ (with corresponding
$\Mstars \la 10^{11.8}\Msun$ and $\Mvir \la 10^{14.6} \Msun$ as shown below). 
This is why the constraints on the VDF evolution become weak toward large 
$\sigma$ as shown in Fig.~\ref{VDFs}. Hence any results from this work outside 
the above ranges must be regarded as extrapolations.

\begin{figure}
\begin{center}
\setlength{\unitlength}{1cm}
\begin{picture}(8,6)(0,0)
\put(-0.6,6.8){\includegraphics{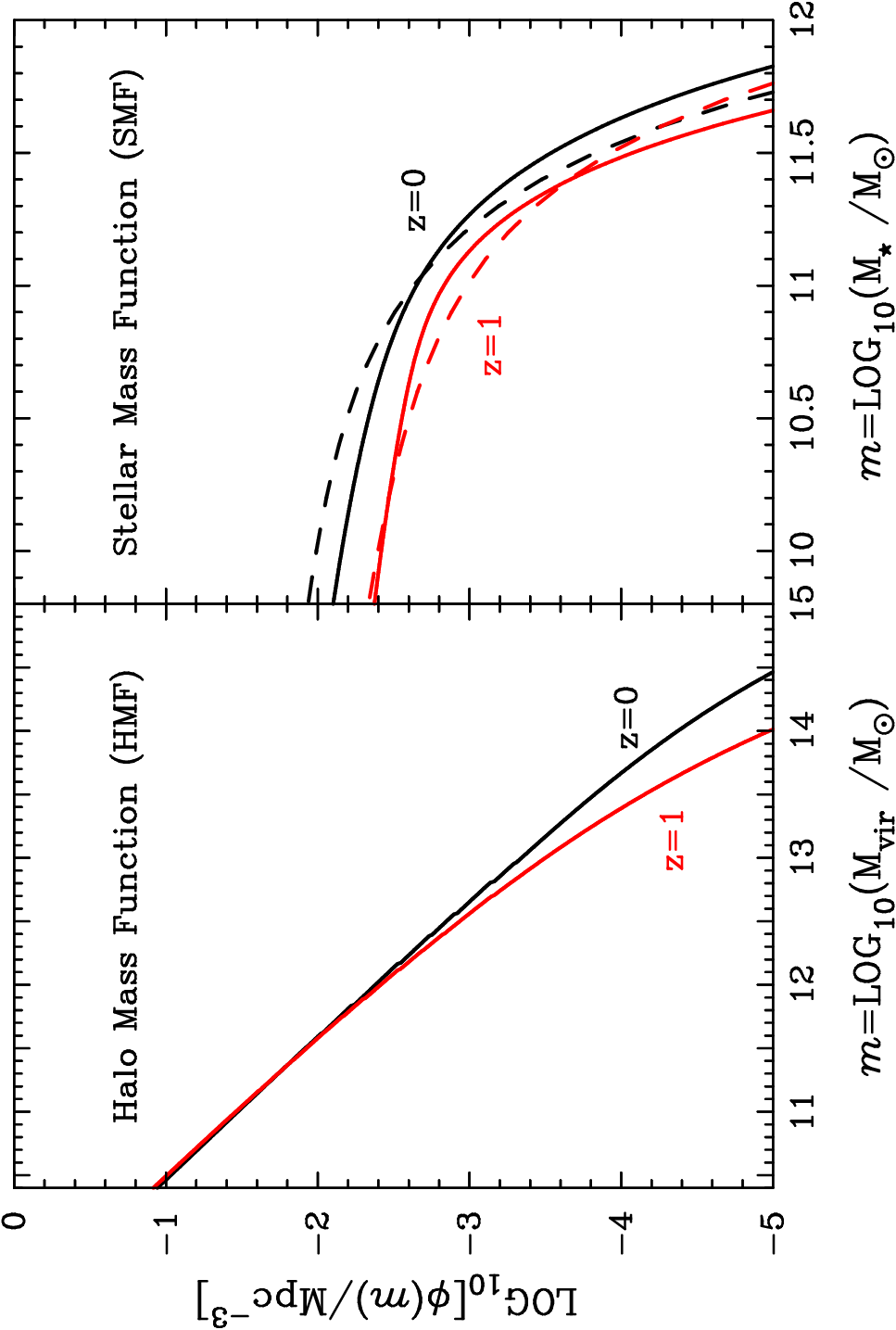}}
\end{picture}
\caption{Left panel: A typical mass function for dark haloes as is produced 
from the $\LCDM$ simulation by \citet{Ree07}. The adopted cosmological 
parameters are  
$(\Omega_{{\rm m}0},\Omega_{{\Lambda}0})=(0.25,0.75)$ and $\sigma_8
=0.8$. The displayed function has been corrected to include
subhaloes (see the texts in Section~3). 
Right panel: Observed stellar mass functions.  Solid curves are 
from the COSMOS survey by \citet{Ilb09} while dashed curves are 
from the Spitzer survey by \citet{Per08}. Notice that the Spitzer
 results show stellar-mass-downsizing evolution of the SMF while 
the COSMOS results do not.
}
\label{MFs}
\end{center}
\end{figure}

For the HMF we use a typical numerical result from N-body simulations under
the current $\LCDM$ cosmology while for the SMF we use the results from some 
representative galaxy surveys. The HMF and the SMF are shown in Fig.~\ref{MFs}
and more details are respectively given in Sections~3 and 4.

\section{Connection between the observational  VDF and the HMF 
from N-body simulations}

In the $\LCDM$ hierarchical structure formation picture the dark halo mass 
function (HMF) evolves over cosmic time as a consequence of hierarchical 
merging (e.g.\ \citealt{WR78,LC93}). Accordingly, the statistical functions of 
galaxies such as the VDF and the SMF are also expected to evolve. However, 
baryon physics
complicates the evolutions of the VDF and the SMF making it non-trivial
to compare the evolutions of the HMF, the SMF, and the VDF one another. 
Conversely, careful analyses of the coevolution of the HMF, the SMF, and
the VDF may reveal key insights into galaxy formation and evolution processes.
Here we compare the evolution of the VDF described in Section~2 with
the evolution of the HMF from cosmological N-body simulations. A comparison 
between the VDF and the SMF is given in the next section.

The HMF may be determined analytically (e.g.\ \citealt{PS74,ST99,ST02}) or 
through $N$-body simulations 
(e.g.\ \citealt{Jen01,Spr05,War06,Ree07,Luk07,Tin08}). 
Recent high resolution $N$-body simulations have determined the HMF reliably 
(e.g.\ \citealt{War06,Ree07,Luk07,Tin08}).  
For the HMF we generate a numerical function using the code provided by 
\citet{Ree07} taking the following cosmological parameters: 
$\Omega_{\rm m0}=0.25$, $\Omega_{\Lambda0}=0.75$ and 
$\sigma_8 =0.8$ consistent with the WMAP5 data (\citealt{Dun09}). 
This function includes only distinct haloes that are not
parts of larger haloes. We correct it to include subhaloes
since they may also host galaxies.\footnote{This correction has only a
relatively minor effect, in particular for large mass.} 
 We use the simulation results by \citet{CWK06} for the number fraction of 
subhaloes ($f_{\rm sub}$) as a function of maximum circular velocity. 
By relating the maximum circular velocity to the halo virial mass using 
the scaling given by \citet{Kly10}, 
we find a varying fraction from $f_{\rm sub} \approx 0.25$ ($\approx 0.18$) at 
$\Mvir \la 10^{11} \Msun$ (this mass scale corresponds to the \citet{CWK06}
completeness limit) to $\approx 0.08$ ($\approx 0.08$) at $\Mvir \ga 10^{13}
\Msun$ for $z=0$ ($z=1$). Notice that the mass of a distinct halo
refers to the epoch under consideration while that of a subhalo is the
mass at the time it accreted onto a larger halo.

The mass of a halo ($\Mvir$) may be linked to the stellar velocity dispersion 
($\sigma$) of the central galaxy for those haloes that host galaxies. 
If the halo did not host a galaxy in its centre, the central velocity dispersion
(of dark matter particles) would be entirely due to the dark mass potential. In
reality, the central galaxy contributes to the central gravitational potential
with the degree of contribution varying from one system to another. 
The functional relation $\sigma$($\Mvir$) will depend not only on the stellar 
mass distribution of the residing galaxy but also how the dark halo has been
modified due to the baryonic physics of galaxy formation. Moreover, the 
stellar mass distribution itself is correlated with $\Mvir$ to some degree.
Thus, we may use the abundance matching relation between $\Mvir$ and $\sigma$ to
gain new insights into the structure of the baryon-modified dark halo and  
the dynamical aspect of galaxy formation and evolution.

\subsection{The $\Mvir$-$\sigma$ relation at $z=0$}

\begin{figure}
\begin{center}
\setlength{\unitlength}{1cm}
\begin{picture}(9,8)(0,0)
\put(-0.7,-2.2){\includegraphics{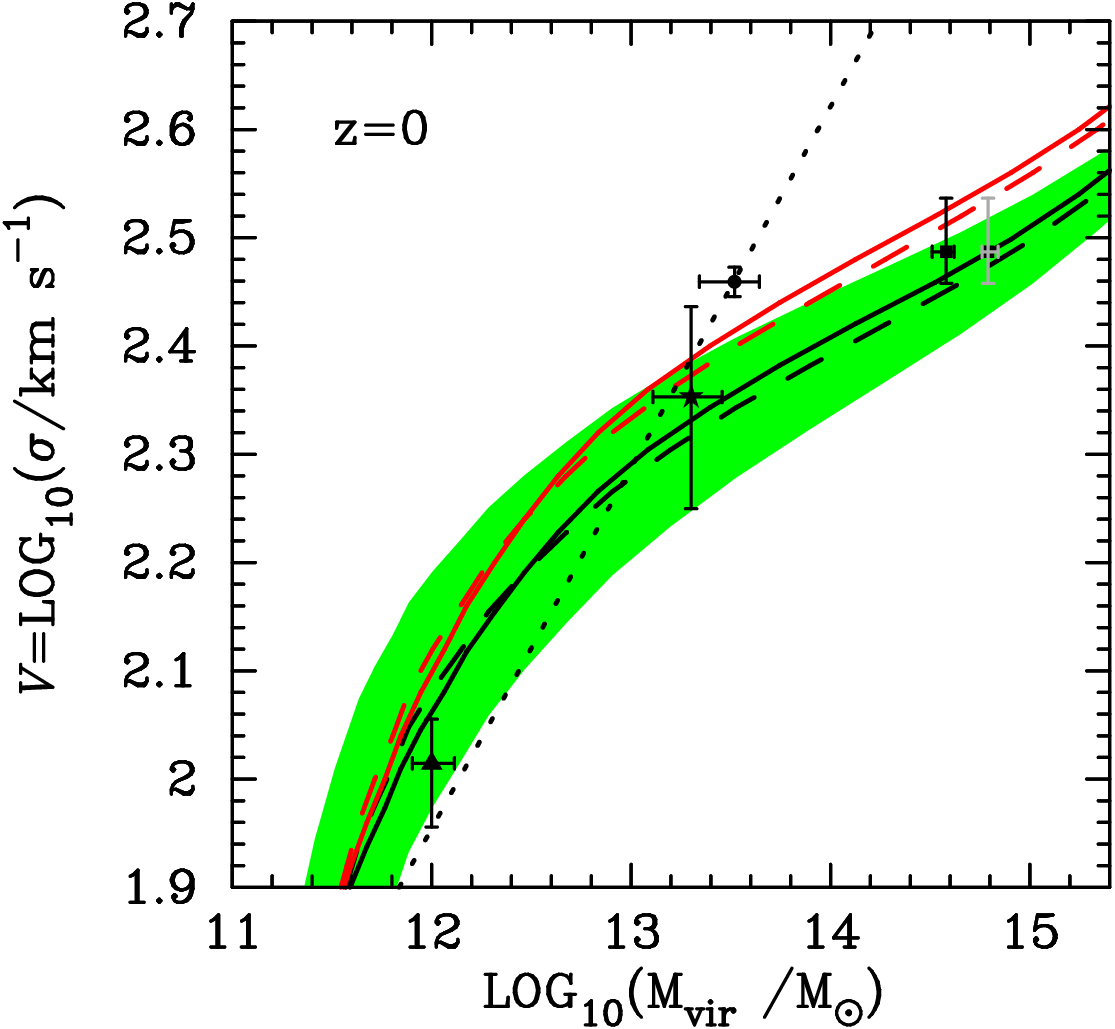}}
\end{picture}
\caption{
The relation between the halo virial mass ($\Mvir$) and the stellar
 velocity dispersion ($\sigma$) of the central galaxy at $z=0$,  
inferred from the abundance matching of the velocity dispersion function and 
the $\LCDM$ halo mass function (see the texts in \S 3). 
 The red curves are the results ignoring the scatter of 
$V \equiv\log_{10} (\sigma/\kms)$ at fixed $\Mvir$. The black curves are 
the results taking into account the scatter shaded green.
See Appendix~A for a brief description of the scatter.
The solid and dashed curves are respectively based on the VDFs by
 \citet{Cha10} and \citet{Ber10}.
The dotted line is a prediction of the SIS model of the halo. The curves are 
compared against the individual measurements for the following systems:
{\it Triangles} - Milky Way (\citealt{Xue08}). 
{\it Star} - A weighted mean of 22 SLACS lenses (\citealt{Gav07}) at 
$z \sim 0.2$ taking the standard deviation of the individual  mean values as 
the error on $\sigma$. {\it Circle} - Lens system Q0957+561 at $z=0.36$. The 
velocity dispersion is from \citet{TF99} while the virial mass is from 
\citet{Nak09}. {\it Squares} - Galaxy cluster Abell 611 (\citealt{New09}).
The right square (gray) is based on the NFW halo model while the left square
(solid) a generalised NFW halo model. 
}
\label{MvirVz0}
\end{center}
\end{figure}

 Fig.~\ref{MvirVz0} shows the abundance matching $\Mvir$-$\sigma$ relation at 
$z=0$.  It shows both the relations ignoring intrinsic scatters 
 and those taking into account an intrinsic scatter distribution of 
$V\equiv \log_{10}(\sigma/\kms)$ as a function of $\Mvir$. For the latter case
the intrinsic scatter distribution is predicted by a bivariate distribution 
of $\sigma$ and $\Mstars$ as a function of $\Mvir$ based on an observed
scatter of $\log_{10}(\Mstars)$ at fixed $\Mvir$ and an observed scatter
distribution of $V$ as a function of $\Mstars$. The reader is referred to
Appendix~A for a brief description and a following work (in preparation) for 
further details.

The abundance matching relation is compared against the measured values of 
$\Mvir$ and $\sigma$ for individual galaxies/clusters with $z \la 0.3$.
Although there are numerous galaxies/clusters for which either $\sigma$ or
$\Mvir$ is reported, only for relatively few systems both $\Mvir$ and $\sigma$ 
have been measured reliably so far. First, we consider the best-studied Milky 
Way galaxy, for which recent measurements appear to be
reasonably concordant (\citealt{KZS02,Bat05,Bat06,Xue08}).\footnote{The 
Andromeda galaxy is also a well-studied example, but the inferred virial 
masses appear to still vary by a factor of 2 (e.g., \citealt{KZS02,SBB08}).}
The data from \citet{Xue08} are displayed in Fig.~\ref{MvirVz0}. 
Second, we display the results for 22 SLACS lensing 
galaxies at mean redshift of $z \sim 0.2$ by \citet{Gav07}. They combine strong
 and weak lensing and stellar kinematics to analyse the systems.  
Third, we consider the first ever discovered lens system Q0957+561 at $z=0.36$,
which is the best studied lens system including a cluster for the lens 
potential. The velocity dispersion for the central galaxy is reported by
\citet{TF99}. The virial mass of the cluster is from \citet{Nak09} who derive
the halo mass through weak lensing and find that their result is consistent 
with the result by \citet{Char02} through X-ray observations. Finally, we
consider galaxy cluster Abell 611 that has been studied extensively by
\citet{New09} through a combination of strong and weak lensing and stellar 
kinematics.  As shown in Fig.~\ref{MvirVz0} these individual measurements
are  in excellent agreement with the $\Mvir$-$\sigma$ relation based on 
the abundance matching of statistical functions. This agreement bolsters the 
validity of the $\Mvir$-$\sigma$ relation at $z=0$. 

Notice that the $\Mvir$-$\sigma$ relation shown in Fig.~\ref{MvirVz0} is a 
curve rather than a straight line. Consequently, it does not match well the 
prediction by the singular isothermal sphere (SIS) halo model (see, e.g.,
 \citealt{LO02}). 
The failure of the SIS model is evident for $\Mvir \ga 10^{13} \Msun$ 
as noticed by several authors (e.g.\ \citealt{LO02,KW01,Blu86}). 
This result confirms the critical
halo mass $M_c \sim 10^{13} \Msun$ below which the baryonic effects start to
 become significant for the inner halo dynamics and structure. 
 However, even for $\Mvir \la 10^{13} \Msun$ the SIS model is not very 
successful in matching the empirically determined $\sigma(\Mvir)$ curve.
This implies that the SIS model is not precise as a `global model' of the
galactic halo despite the fact that a range of observational constraints
support the isothermal profile for the inner part of the halo (see 
\citealt{Cha10} and references theirin). The underprediction of $\sigma$ by 
the SIS model for $\Mvir \la 10^{13} \Msun$ probably reflects
the neglected concentration of the  halo.  
The curvature in the $\Mvir$-$\sigma$ relation may reflect 
 the systematic variation of halo concentration but
may also imply the varying baryonic effects on the halo structures.
The internal structures of the haloes may be constrained by combining dynamical 
constraints with the empirical $\Mvir$-$\sigma$ relation (in preparation).

\subsection{The coevolution of the HMF and the VDF}

\begin{figure}
\begin{center}
\setlength{\unitlength}{1cm}
\begin{picture}(8,6)(0,0)
\put(-0.9,6.9){\includegraphics{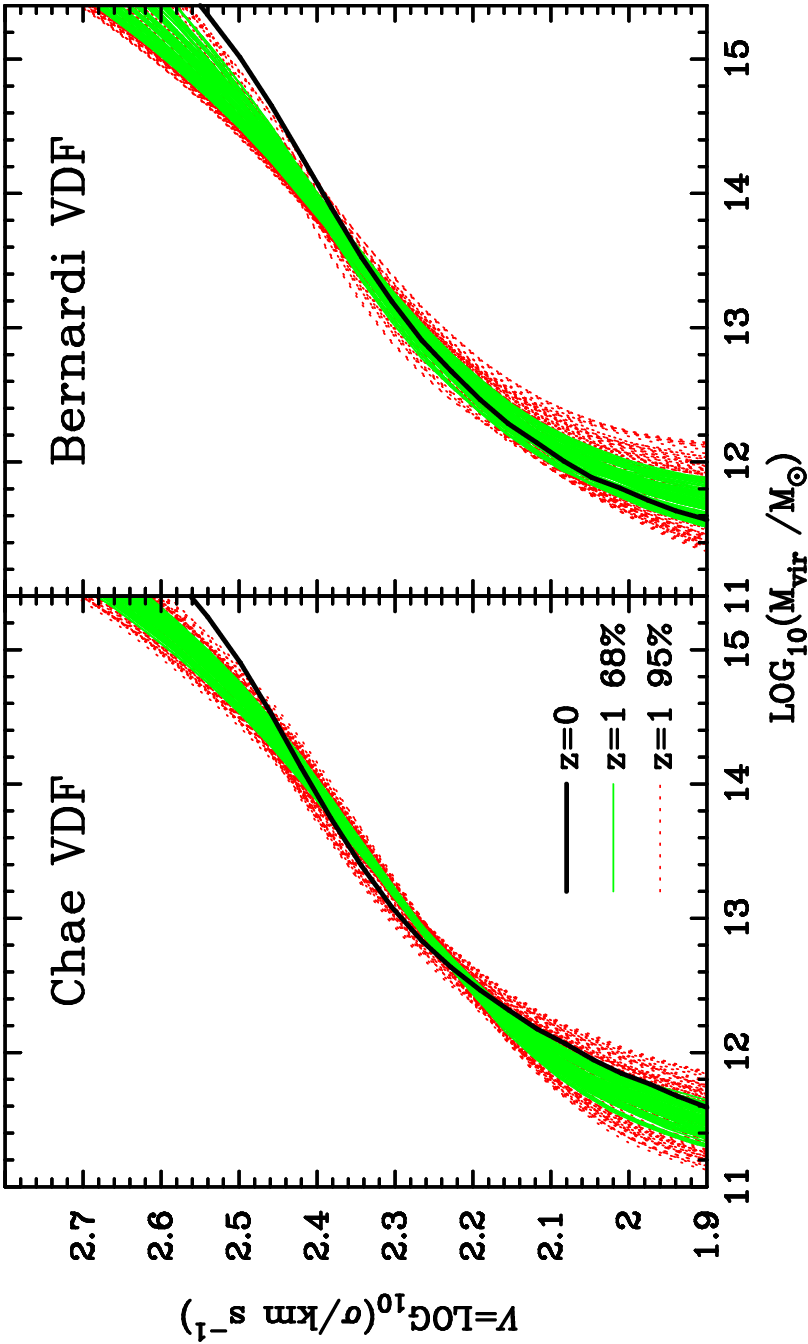}}
\end{picture}
\caption{
The abundance matching $\Mvir$-$\sigma$ relation at $z=1$ is compared with 
that at $z=0$. It is consistent with zero evolution between $z=1$ and $z=0$
for the strong lensing probed range $\sigma \la 300\kms$ 
($\Mvir \la 10^{14.6} \Msun$; see \S 2). 
}
\label{MvirV}
\end{center}
\end{figure}

In Fig.~\ref{MvirV} we compare the abundance matching $\Mvir$-$\sigma$ relations
at $z=1$ and $0$. This comparison shows little sign of evolution in the 
$\Mvir$-$\sigma$ relation for the strong lensing probed range 
$\sigma \la 300\kms$ ($\Mvir \la 10^{14.6} \Msun$; see \S 2).
The near constancy in the $\Mvir$-$\sigma$ relation with $z$ implies that 
the HMF and the VDF are coevolving in parallel. Namely,
as the halo grows in mass over cosmic time, the central stellar velocity 
dispersion grows in accordance. The natural question to ask is then what the 
origin of this coevolution is. We discuss this in \S 5.

\begin{figure}
\begin{center}
\setlength{\unitlength}{1cm}
\begin{picture}(8,7)(0,0)
\put(-0.7,6.8){\includegraphics{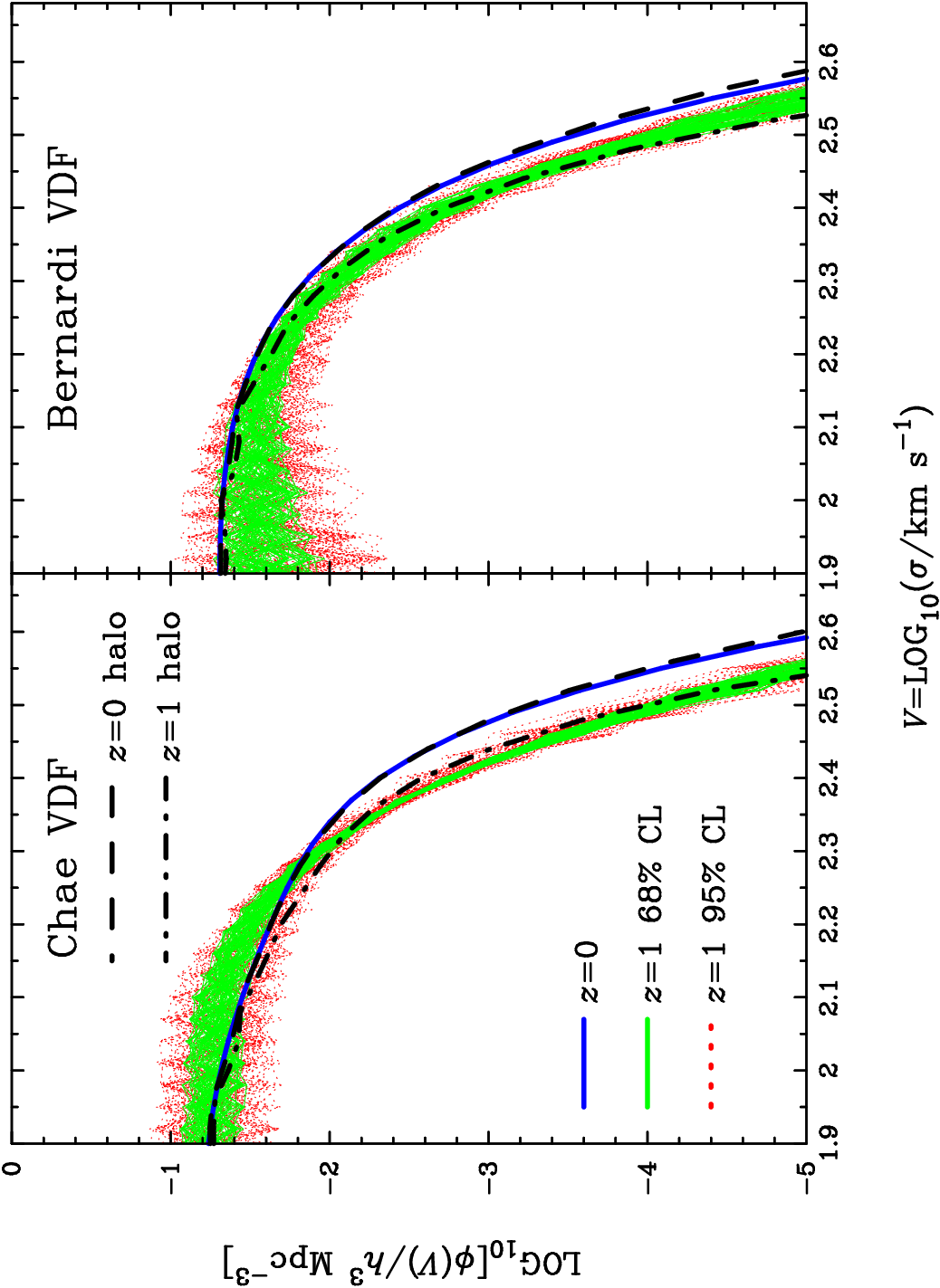}}
\end{picture}
\caption{
The observationally constrained VDFs at $z=0$ and $z=1$ are compared with 
the VDFs predicted from the $\LCDM$ halo mass function using the empirically 
determined relation $\sigma(\Mvir)$ at $z=0$ assuming zero evolution 
(see Fig.~\ref{MvirV}). There is a reasonably good match between the 
observationally derived VDF evolution and the halo-predicted VDF evolution.
}
\label{VDFhalo}
\end{center}
\end{figure}

Alternatively, we may transform the HMF into a VDF using the $z=0$ relation of 
Fig.~\ref{MvirV} and assuming a certain evolution of the $\Mvir$-$\sigma$ 
relation. Fig.~\ref{VDFhalo} shows the VDFs predicted from the HMFs assuming 
zero evolution of the $\Mvir$-$\sigma$ relation. 
 In Fig.~\ref{VDFhalo} the HMF-converted VDFs are compared with the 
observationally derived local VDFs and the lensing constrained VDFs at $z=1$. 
The VDFs at $z=0$ are in excellent agreement with each other. 
The VDFs at $z=1$ are also in agreement with each other.
This exercise shows that under the simple assumption of the constancy of
the $\Mvir$-$\sigma$ relation in time, the evolution of the HMF 
predicted by the current $\LCDM$ cosmology can match well the evolution of 
the VDF constrained by strong lensing statistics for $0 \la z \la 1$.

\section{Connection between the observational VDF and the SMF 
from galaxy surveys}

Many recent surveys of galaxies have been used to constrain the evolution of
galaxies through the LF or/and the SMF. The results are at variance. 
 Many results argue for relatively little evolution in the number
 density of most massive galaxies and greater evolution of less massive
galaxies over cosmic time, i.e.\ a ``stellar mass-downsizing'' 
(anti-hierarchical) behaviour 
(e.g., \citealt{Cim06,Fon06,Poz07,Con07,Sca07,Coo08,Per08,Mar09}),
although there are results that do not particularly support a 
 mass-downsizing evolution (e.g., \citealt{Bel04,Fab07,Bro08,Ilb09}). 
 The variance for the evolution of the SMF is not well understood but may be 
due to errors in measurements and modelling of the SMF (see \citealt{LS09}) 
and galaxy sample biases caused by cosmic variance 
(see, e.g., \citealt{Fab07,Cat08,Str08} for discussions). 

We have seen in the previous section that the strong lensing constrained VDF
evolution is in line with the theoretical HMF evolution. How well would the VDF 
match the observed SMF from galaxy surveys? 
What would be the correlation between stellar mass 
($\Mstars$) and velocity dispersion ($\sigma$) and its evolution up to $z=1$?
The coevolution of the SMF and the VDF will depend on the evolution of the 
$\Mstars$-$\sigma$ relation. 
Hence the evolution of the VDF can be tested against the evolution of the SMF 
only when the $\Mstars$-$\sigma$ relation is known (or assumed) as a function 
of cosmic time. Conversely, by matching the observed SMF evolution from galaxy
surveys with the VDF evolution from strong lensing statistics we may infer the 
evolution of the $\Mstars$-$\sigma$ relation.  In the following we take
the latter approach.

  We match the VDFs by \citet{Cha10} and 
\citet{Ber10} (their evolutions being constrained by strong lensing) with
 two total SMFs from galaxy surveys (see Fig.~\ref{MFs}) that are 
qualitatively different and are intended to encompass the current range of 
observations.
One is the COSMOS SMF by \citet{Ilb09} measured using 192,000 galaxies from the
 COSMOS 2-deg$^2$ field. The COSMOS sample size is much larger than any other 
single data set that has been used to derive the SMF up to $z \ga 1$. 
For the \citet{Ilb09} SMF $z=0$ function is actually for $0.2 < z < 0.4$ while
 $z=1$ function is a mean of $0.8 < z < 1.0$ and $1.0 < z < 1.2$ functions. 
Notice that the \citet{Ilb09} SMF does not show a stellar mass-downsizing 
evolution. 
The other is the Spitzer SMF by \citet{Per08}.\footnote{\citet{Per08} 
adopt the Salpeter IMF to calculate their stellar masses. To convert their 
stellar masses to those based on the Chabrier IMF, we divide by 1.7.} 
This is a typical SMF that shows a downsizing behaviour.
For the \citet{Per08} SMF $z=0$ function is actually for $0 < z < 0.2$ while
 $z=1$ function is a mean of $0.8 < z < 1.0$ and $1.0 < z < 1.3$ functions. 
We note that the \citet{Per08} sample covers a sky area of only 
$\sim 664$~arcmin$^2$ and contains $\sim 28,000$ sources for $0 < z < 4$.

\subsection{The $\Mstars$-$\sigma$ relation at $z=0$}

\begin{figure}
\begin{center}
\setlength{\unitlength}{1cm}
\begin{picture}(8,8)(0,0)
\put(-0.7,-2.1){\includegraphics{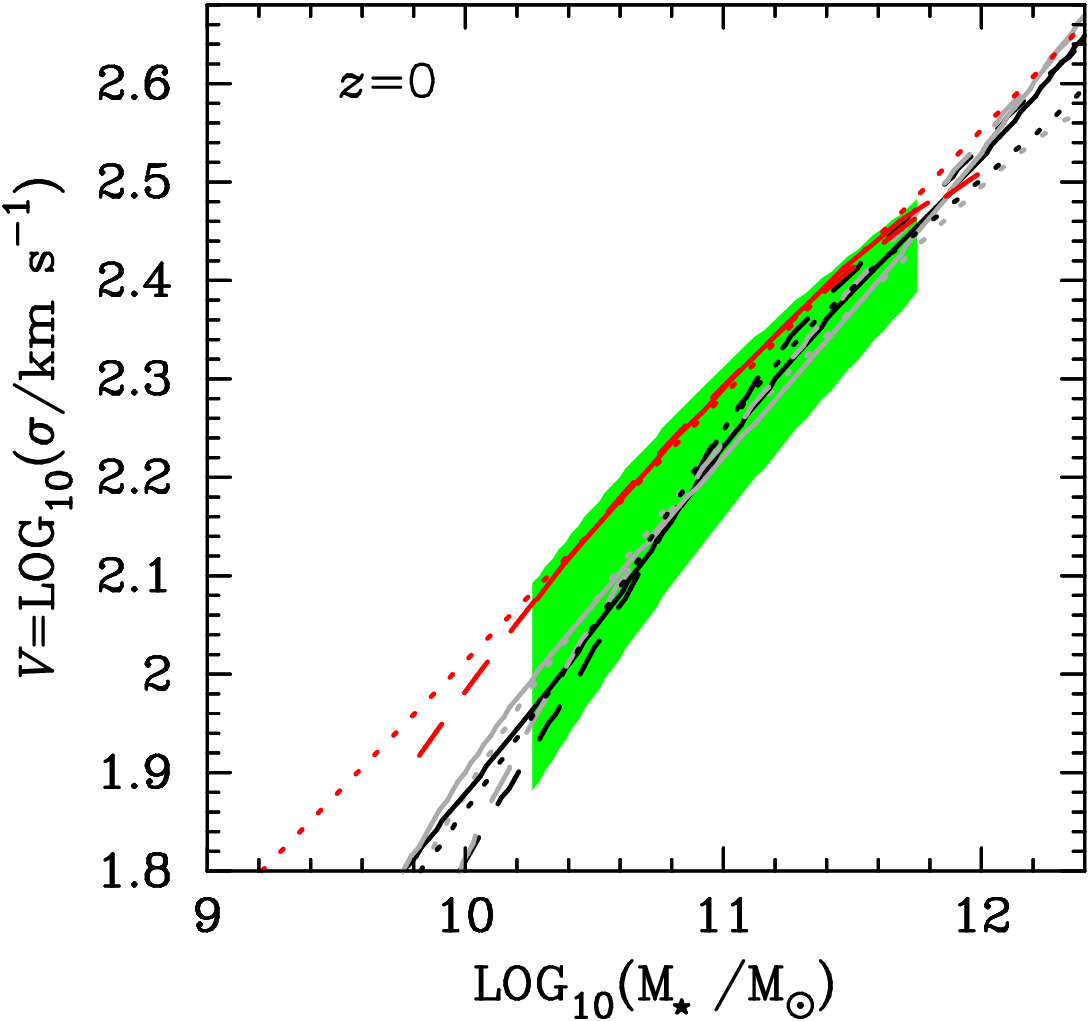}}
\end{picture}
\caption{
The abundance matching relation between the stellar mass ($\Mstars$) and the 
stellar velocity dispersion ($\sigma$) of galaxies at $z=0$. 
Black (gray) solid, dashed, and dotted curves are respectively the results 
of matching the \citet{Cha10} (\citealt{Ber10}) VDF with the COSMOS
SMF by \citet{Ilb09}, the Spitzer SMF by \citet{Per08}, and the SDSS SMF
by \citet{Ber10} for all galaxies. These abundance matching results have been
corrected for the effects of the intrinsic scatter shaded green. The intrinsic
scatter and its effects are described in Appendix~A. The red solid, dashed and
dotted curves/lines are the measured median  $\Mstars$-$\sigma$ relations
respectively by \citet{Des07}, \citet{HB09}, and \citet{Sha10} for early-type
galaxies. Notice that the abundance matching relations agree well with
the measured early-type relations at high $\Mstars$ but deviate systematically
as $\Mstars$ gets lower because the late-type contribution becomes increasingly 
larger.
}
\label{MstarVz0}
\end{center}
\end{figure}

Fig.~\ref{MstarVz0} shows several examples of the abundance matching relation 
between $\Mstars$ and $\sigma$ at $z=0$. For $z=0$ only we consider the 
\citet{Ber10} SMF as well as the COSMOS and the Spitzer SMFs. These results
have been obtained taking into account the effect of an adopted intrinsic 
scatter (the region shaded green) for $V \equiv \log_{10}(\sigma/ \kms)$ of 
$0.115-0.039\times(m-10)$ with $m \equiv \log_{10}(\Mstars/\Msun)$ from 
\citet{Des07} (see Appendix~A). The details on the effect of the intrinsic 
scatter can be found in Appendix~A. Notice that the abundance matching 
$\Mstars$-$\sigma$ relations have mild curvatures.
For a linear approximation $V= b m + {\mbox{const}}$, the slope $b$ varies
from $b=[0.23,0.34]$ for $m > 11.5$ to $b=[0.35,0.49]$ for $m < 10.5$.
The abundance matching relations for all galaxies are compared with the 
directly measured median relations for early-type galaxies in the literature
(\citealt{Des07,HB09,Sha10}). The abundance matching relations agree well with
the early-type relations for $\Mstars \ga 10^{11.6} \Msun$. However, 
as $\Mstars$ decreases, the abundance matching relations deviate 
systematically and increasingly from the early-type relations. 
This is expected and can be well understood by the fact 
that the late-type relation is different from the early-type 
relation and the late-type contribution to the total relation increases as 
$\Mstars$ decreases.  

\subsection{The $\Mstars$-$\sigma$ relation at $z=1$ and its 
evolution to $z=0$}

In Fig.~\ref{MstarV}, the abundance matching $\Mstars$-$\sigma$ relation at 
$z=1$ is shown and compared with that at $z=0$. 
The relation at $z=1$ is also compared against the individual 
data points for $0.7 < z < 1.3$ from \citet{dSA05} and \citet{vdW05}.
The resulting evolution of the $\Mstars$-$\sigma$ relation varies 
depending mostly on the adopted SMF. Notice that the strictly valid range 
probed by the data is $1.97 \la V[\equiv \log_{10}(\sigma/\kms)] \la 2.47$ 
corresponding to $10.2 \la m[\equiv \log_{10}(\Mstars/\Msun)]  \la 11.8$ 
(see \S 2). The results outside this range are extrapolations.

\begin{figure}
\begin{center}
\setlength{\unitlength}{1cm}
\begin{picture}(9,9)(0,0)
\put(-0.5,-1.6){\includegraphics{f7.eps}}
\end{picture}
\caption{
The $\Mstars$-$\sigma$ relation at $z=1$ is compared with that at $z=0$, 
inferred from the abundance matching of the strong lensing constrained 
evolutions of the VDFs of \citet{Cha10} and \citet{Ber10} with the observed 
SMFs from galaxy surveys (see the texts in \S 4). Two SMFs are used: the 
COSMOS SMF by \citet{Ilb09} and the Spitzer SMF by \citet{Per08}. 
The blue data points
are the weighted means and their errors (thin error bars are dispersions) for 
mass intervals of 0.4 dex based on 47 galaxies for $0.7 < z < 1.3$ from 
\citet{dSA05} and \citet{vdW05}. All stellar masses are for the Chabrier IMF.
} 
\label{MstarV}
\end{center}
\end{figure}

For the COSMOS SMF the $\Mstars$-$\sigma$ relation is consistent with zero 
evolution. The relation at $z=1$ is also consistent with the measured data 
points.
On the other hand, for the Spitzer SMF (a typical downsizing SMF) the $z=1$ 
relation deviates systematically from the $z=0$ relation in particular for 
massive galaxies with $\Mstars \ga 10^{11} \Msun$. 
 This appears to be the case for any downsizing SMF. This is because the
evolutionary behaviour of a downsizing SMF is dissimilar to that of the VDF.
 This implies that a downsizing SMF requires a differential evolution in
the $\Mstars$-$\sigma$ relation. In particular, according to the
downsizing SMF $\sigma$ has to be lower at $z=1$ than $z=0$
at fixed $\Mstars$ ($\ga 10^{11.2-11.4} \Msun$).

Then, which of the above cases (the non-evolving or the evolving case) of the 
$\Mstars$-$\sigma$ relation would be more consistent with other independent 
 results on the structural evolutions of galaxies?

\subsubsection{Comparison with observed structural evolutions of galaxies}

\begin{figure}
\begin{center}
\setlength{\unitlength}{1cm}
\begin{picture}(9,11)(0,0)
\put(-0.8,-0.6){\includegraphics{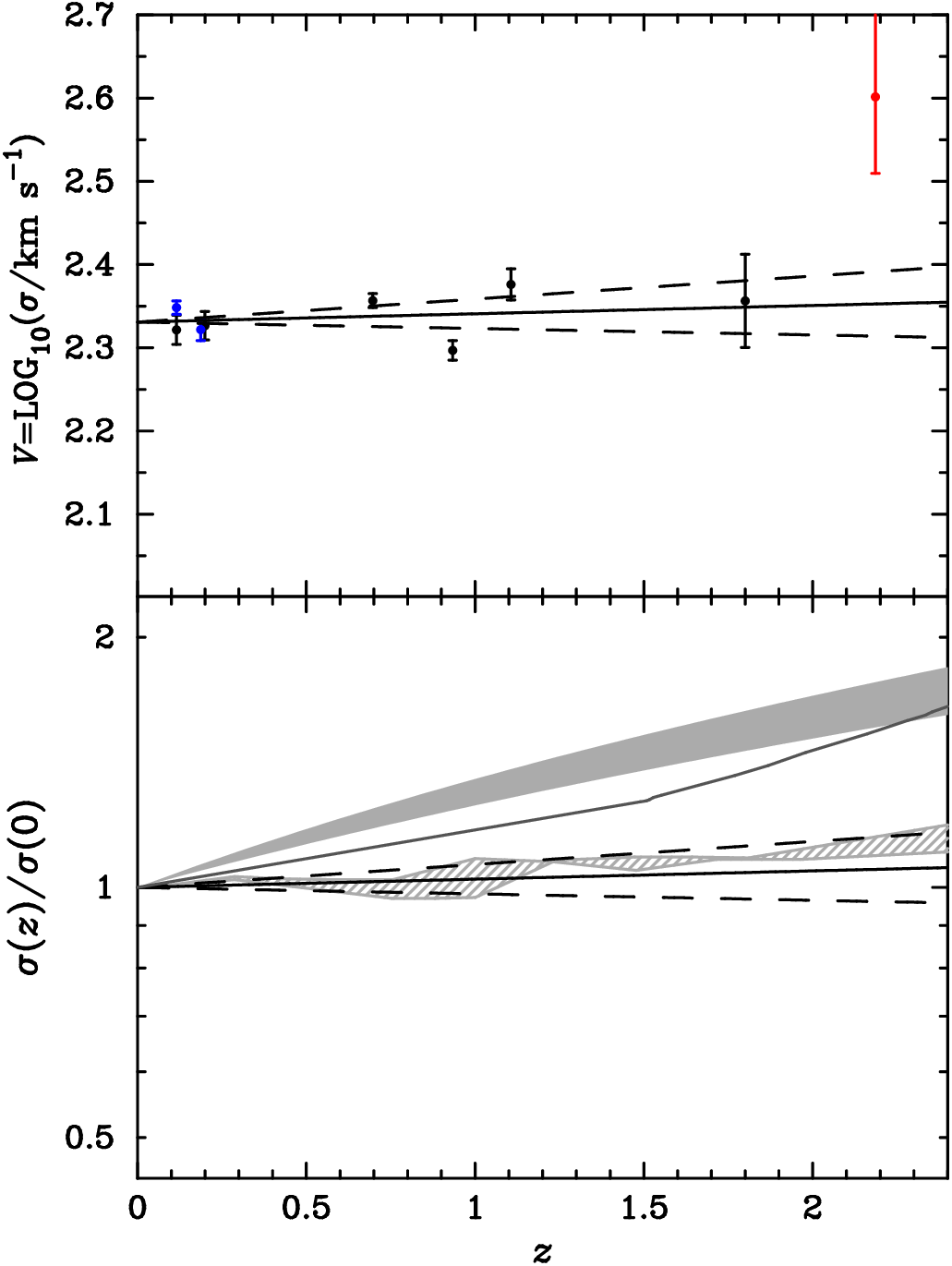}}
\end{picture}
\caption{
{\it Upper panel}:
The observed stellar velocity dispersion ($\sigma$) as a function of $z$ at 
fixed stellar mass $\Mstars=10^{11} \Msun$. The data points are based on  
galaxies with measured stellar masses in the range 
$10.75 < m[\equiv {\rm log}_{10} (\Mstars/\Msun)] < 11.25$. 
Furthermore, to estimate the velocity dispersion at $m=11$ as precisely as 
possible we use an empirical relation of $V = b m + {\rm const}$ where 
$V = {\rm log}_{10}(\sigma/ \kms)$ and we take $b=0.34$ as estimated from 
Fig.~\ref{MstarVz0} for $10.75 < m < 11.25$ (the results are insensitive to
the exact value of $b$ for $b > 0.2$).  The references for the data are as 
follows: (1) $z < 0.15$, 8 (or 17 for blue point) galaxies -- 
 \citet{Bol08} \& \citet{Gri09}   (or \citet{Aug09})
(2) $ 0.15 \le z < 0.4$, 9 (or 11 for blue point) galaxies -- 
 \citet{Bol08} \& \citet{Gri09} 
 (or \citet{Aug09})
(3) $ 0.6 \le z < 0.8$, 7 galaxies -- \citet{vdW05}  
(4) $ 0.8 \le z < 1.0$, 10 galaxies -- \citet{vdW05}, \citet{dSA05}  
(5) $ 1.0 \le z < 1.3$, 8 galaxies -- \citet{vdW05}, \citet{dSA05} 
(6) $ 1.6 \le z < 1.8$, 7 galaxies -- \citet{Cap09},
(7) z=2.186, 1 galaxy (red point) -- \citet{vanD09}.  The solid line is the 
best-fit in the least-square fit of the data points and the dashed lines 
represent the errors in the slope. The single galaxy at z=2.186 is not used for
 the fit but consistent with the fit result at the $2\sigma$ level.
 {\it Lower panel}:
The evolution factor of the velocity dispersion as a function of $z$. 
The black solid and dashed lines are the fit results from the upper panel.
The light hatched area is the prediction by \citet{Hop09b}. The light gray area
is the result based on the \citet{Hop09b} model presented by \citet{CT09} who 
use local SDSS data and similar data for $z>0.5$ as used here but without 
converting the measured velocity dispersions to the values at a fixed stellar 
mass (they use a broad range 
$0.5 \times 10^{11} \Msun \la \Mstars \la 2 \times 10^{11} \Msun$). 
The dark gray solid curve is the prediction by \citet{Hop10}.
}
\label{Vzevol}
\end{center}
\end{figure}

According to recent studies on the structural evolutions of galaxies,
there are observational indications that galaxy size evolves at fixed
stellar mass (e.g.\ \citealt{Tru07,vdW08,Cim08,vanD08,Bez09}). However, 
more recent studies find that physical mass densities (as opposed to effective 
densities) evolve little (\citealt{Hop09,Bez09}).
This implies that velocity dispersion might evolve little at fixed stellar mass.
Indeed, \citet{CT09} find a slow evolution of $\sigma(\Mstars)$  
since $z \sim 1.6$ from an analysis of spectra of massive galaxies
(see also \citealt{Cap09}). Nevertheless, \citet{CT09} find a higher $\sigma$ 
at a higher $z$ for 
$0.5 \times 10^{11} \Msun \la \Mstars \la 2 \times 10^{11} \Msun $ (an evolution
from $\sigma \sim 180 \kms$ at $z \sim 0$ to $\sim 240 \kms$ at $1.6$).
However, as shown in Fig.~\ref{Vzevol} a similar data set actually appears to 
indicate no evolution at all. The velocity dispersion at a fixed stellar mass
 of $\Mstars=10^{11}\Msun$ rather than a range stays constant at 
$\sigma \approx 210 \kms$ between $z \sim 0$ and $\sim 1.8$.
Notice that we only use galaxies with 
$10.75 < {\rm log}_{10} (\Mstars/\Msun)] < 11.25$ and we convert the measured 
value of $\sigma$ at the measured value of $\Mstars$ for each galaxy to that at
$\Mstars=10^{11}\Msun$ using an empirical relation found in Fig.~\ref{MstarVz0}.
This prescription largely removes any systematic error arising from the 
differences in the stellar masses of the galaxies in the different redshift 
bins. Another difference between the \citet{CT09} analysis and ours
is that for local galaxies \citet{CT09} use SDSS DR6 data to derive stellar 
masses while we use only SLACS galaxies for which two independent stellar mass 
measurements are available based on SDSS (\citealt{Gri09}) and HST 
(\citealt{Aug09}) photometric data. 

Incidentally, all the abundance matching 
results shown in Fig.~\ref{MstarV} imply no or little evolution of $\sigma(z)$
at $\Mstars=10^{11}\Msun$ regardless of the VDF or the SMF used. Hence,
our abundance matching results are in excellent agreement with, but at the
same time are not distinguished by, our analysis of
 of the data from the literature for individual galaxies for $0 \la z \la 2$.
However, it is important to notice that none of the current observational
results on the structural evolutions of galaxies indicate a negative
evolution in  $\sigma$ with $z$ for any $\Mstars$. Observational
indications (e.g.\ stellar mass densities) are such that velocity
dispersions cannot be lower at a higher $z$ in case of evolution. 
Hence, according to our abundance matching
results the downsizing SMF is inconsistent with the VDF evolution from
strong lensing by \citet{Cha10} because it requires a
lower $\sigma$ at a higher $z$ for massive galaxies.

\subsubsection{Comparison with predictions from cosmological hydrodynamic simulations}

At fixed stellar mass cosmological hydrodynamic simulations from the literature 
also find slow or little evolutions of velocity dispersion. For example,
\citet{Hop09b} combine dark halo merging with hydrodynamic 
simulation results and observed empirical properties of galaxies to find little
evolutions of $\sigma(z)$ at  fixed $\Mstars$ (Fig.~\ref{Vzevol}).
 \citet{Hop10} find slow evolutions through more realistic cosmological 
simulations taking into account various effects including equal and minor 
merging, adiabatic expansion and observational effects (Fig.~\ref{Vzevol}).
\citet{CT09} predict based on the \citet{Hop09b} model a somewhat greater 
evolution using their analysis of spheroid size evolutions (Fig.~\ref{Vzevol}). 
 These simulations are broadly consistent with the 
constraints from the current data as analysed above (Fig.~\ref{Vzevol}) and
the abundance matching results (Fig.~\ref{MstarV}) in the sense that the 
predicted evolutions are slow and can be made in principle to agree with the 
observational constraints. 
Furthermore, \citet{Hop09b}  find that the evolution of
$\sigma(z)$ at fixed $\Mstars$ has little sensitivity on $\Mstars$ for
$10^9 \Msun \le \Mstars \le 10^{12} \Msun$. These simulation results
are consistent with our abundance matching results based on
the non-downsizing SMF but not with the downsizing SMF.

To sum up, our abundance matching results based on the non-downsizing SMF
are broadly consistent with observational constraints on the
structural evolutions of galaxies and cosmological hydrodynamic 
simulation results. However, the downsizing SMF gives a differential
evolution in $\sigma(\Mstars)$ with $z$ that would not be consistent with
observational constraints or simulation results.

\section{Implications for  the $\LCDM$ paradigm and galaxy evolution}

At the heart of the current hierarchical structure formation theory is the
bottom-up build-up of dark matter haloes. 
Given that galaxies are believed to be born and centered in those haloes, 
what would be the evolutionary patterns of galaxies? Unlike dark haloes, 
 galaxies have two distinctive properties, namely, the photometric property
and the dynamical property. Hence, there are two evolutionary properties
of galaxies to be considered. The star formation history of galaxies
gives rise to above all the evolutionary patterns in the luminosity and stellar
mass functions of galaxies. Most cosmological observations have been devoted to
 the photometric properties. Galaxy formation models, whether semianalytical 
or hydrodynamical, have also tried to reproduce the photometric properties 
rich in observational data. Notice that dark haloes do not have such 
photometric properties. This means that connections between the photometric 
properties of galaxies and dark haloes are indirect and challenging.
Galaxy evolution in its dynamical property, which is the focus of this work, 
may be characterized by the evolutionary patterns in the velocity (dispersion) 
functions of galaxies. 
One may wonder whether the dynamical property of galaxies may be more 
intimately linked to dark haloes than the photometric properties do. 
How is the dynamical property of galaxy evolution related to dark haloes?
What would be the role of the photometric property of galaxy evolution in this
context?

Current galaxy formation models cannot 
yet reliably predict central dynamical properties of galaxies. 
In this work we have compared the lensing-constrained evolution of the
VDF with the $\LCDM$ predicted evolution of the HMF and the observed evolution
of the SMF. The main result is that the halo virial mass ($\Mvir$), the galaxy
stellar mass ($\Mstars$) and the central line-of-sight stellar velocity
dispersion ($\sigma$) are positively coevolving for the
probed redshift range of $0 \la z \la 1$. 
 What are the implications of the results from this work for galaxy formation 
and evolution under the $\LCDM$ hierarchical paradigm?

{\it Halo mass-velocity dispersion relation:} We find that 
$M_{\rm vir}$-$\sigma$ relation does not evolve between $z=1$ and $z=0$ 
(Fig.~\ref{MvirV}) for the entire probed range of halo mass from 
galactic haloes to cluster haloes. This empirical finding is insensitive
to the choice of the HMF and the VDF from current simulations and observations
under a concordance $\LCDM$ cosmological model.

Pure dark halo simulations predict that a halo of mass
$M_{\rm vir}$ at $z=1$ is smaller (i.e.\ smaller $R_{\rm vir}$) but less 
concentrated (i.e.\ smaller $c_{\rm vir}$) than that at $z=0$.
As to the central velocity dispersion the two effects are opposite so that
we may expect $\sigma$($M_{\rm vir}$) to evolve little as far as pure haloes
are concerned. Let us consider this quantitatively using a simple model.
Without dissipational galaxy formation, the evolution of
the central velocity dispersion would be primarily determined by the evolutions
of the virial mass ($\Mvir$), the virial radius ($\rvir$) and the concentration 
($\cvir$). The velocity dispersion is expected to increase (decrease, increase)
if $\Mvir$ ($\rvir$, $\cvir$) increases while the other two parameters are held 
constant. Cosmological $N$-body simulations predict that all three parameters
(i.e.\ $\Mvir$, $\rvir$ and $\cvir$) increase as cosmic time evolves forward.
Suppose $\sigma = \vvir f(\cvir,r/\rvir)$ where $\sigma$ is the velocity
dispersion in the central region (e.g.\ within $0.01\rvir$), 
$\vvir=\sqrt{G \Mvir/\rvir}$ is the circular velocity at the virial radius,
and $f(\cvir,r/\rvir)$ is a model-dependent factor relating the two.
$N$-body simulations show that $\Mvir$, $\rvir$ and $\cvir$ all increase
roughly by a factor of $2$ from $z=1$ to $0$ (see, e.g., \citealt{Wec02}).
Then, $\vvir$ stays roughly constant and $f(\cvir,r/\rvir)$ increases by
about 15\% from $z=1$ to $0$ for an isotropic NFW model (see \citealt{LM01}). 
Hence we expect some enhancement in the velocity dispersion
(i.e.\ a positive coevolution) in the course
of the hierarchical growth of a pure dark halo from $z=1$ to $0$.

For realistic haloes hosting (dissipationally formed) galaxies
hydrodynamic simulations can be used to predict the evolution of 
$\sigma$($\Mvir$). Unfortunately, current hydrodynamic simulations do not 
predict robustly the baryonic effects on halo structures 
(e.g.\ \citealt{Blu86,Gne04,Aba09,Tis10,Fel10}). Specifically,
recent hydrodynamic simulations  overpredict
$\sigma$ at a given $\Mvir$ (e.g.\ \citealt{Tis10,Fel10}).

The finding that the $\Mvir$-$\sigma$ relation does not evolve for 
$0 \le z \le 1$ offers new insights into galaxy formation and evolution. 
It implies that the dynamical property of the central galaxy of a halo has 
little to do with its history but is dictated by the final halo
virial mass at least since $z=1$. Remarkably, this is the case for all haloes 
probed (with $\sigma \ga 100 \kms$). 
 Implications of this finding are discussed
below in the context of the coevolution of $\Mvir$, $\sigma$ and $\Mstars$.

{\it Stellar mass-velocity dispersion relation:}
The $\Mstars$-$\sigma$ relation at $z=0$ shows a power-law relation  
$\Mstars \propto \sigma^{\gamma_{\rm SM}}$ with a varying power-law index 
$\gamma_{\rm SM}$ ranging from [2.9, 4.4] for $\Mstars > 10^{11.5} \Msun$ 
to [2.0,2.9] for $\Mstars < 10^{10.5} \Msun$.
 Let us compare the  $\Mstars$-$\sigma$ relation with 
power-law correlations between luminosity and internal velocity parameter, 
namely the Tully-Fisher relation for the late-type population and 
the Faber-Jackson relation for the early-type population. 
The observed Tully-Fisher relation exponent $\gamma_{\rm TF}$ lies between 2.5 
and 3.5 (see \S 2.3 or \citealt{Piz07}). The traditional value for the 
Faber-Jackson exponent $\gamma_{\rm FJ}$ 
for early-type galaxies is $\approx 4$. However, an extensive
analysis of SDSS DR5 early-type galaxies reveals that $\gamma_{\rm FJ}$ varies 
systematically from  $2.7 \pm 0.2$ at $L_*$ to $4.6 \pm 0.4$ at the upper 
luminosity end (\citealt{Cho07}; see also \citealt{Des07}). 
The abundance matching $\Mstars$-$\sigma$ relation for all galaxies
 can match well these Faber-Jackson/Tully-Fisher relations
in conjunction with measured $\Mstars/L$ ratios (e.g.\ \citealt{Bel03}).

In Fig.~\ref{MstarV} the $\Mstars$-$\sigma$ relation at $z=1$ is 
compared with that at $z=0$ based on two VDFs and two SMFs that
are meant to encompass the current range of observations.
As can be seen in the figure, the implied evolution depends sensitively on
the adopted SMF and to a less degree on the adopted VDF. 
The relation based on the COSMOS SMF is consistent with
zero evolution in $\sigma(\Mstars)$ between $z=1$ and $0$.
On the other hand, the relation  based on the Spitzer SMF (a typical downsizing
SMF) implies a differential evolution in $\sigma(\Mstars)$: for 
$\Mstars \ga  10^{11} M_{\odot}$  the implied evolution in $\sigma$ with
redshift at fixed $\Mstars$ is negative while it is positive for
$\Mstars \la  10^{11} M_{\odot}$. This means that based on the 
downsizing SMF a galaxy at $z=1$ would have a shallower (steeper) mass profile 
than the local counterpart of the same stellar mass for 
$\Mstars \ga 10^{11} M_{\odot}$ ($\Mstars \la 10^{11} M_{\odot}$). 

How the above results on the evolution in $\sigma(\Mstars)$ are compared 
with other independent results on the structural evolutions of galaxies? 
First of all, we find little evolution in $\sigma(z)$ at $\Mstars=10^{11}\Msun$
for $0 \la z \la 1.8$ from a careful analysis of the data in the literature
(see Fig.~\ref{Vzevol}). This is in excellent agreement with the above abundance
matching results. However, it cannot unfortunately distinguish the abundance
matching results because $\Mstars= 10^{11} \Msun$ happens to be the critical
mass for the downsizing SMF at which the evolution changes the sign.
Second, many observational studies find a negative size evolution of galaxies 
with redshift implying a more steeply declining stellar mass profile at a higher
$z$ (e.g.\ \citealt{Tru07,vdW08,Cim08,vanD08}). 
However, more recent studies find that stellar mass density profiles of the
inner regions up to several kilo-parsecs are consistent with no evolution for
massive galaxies with $\Mstars \ga  10^{11} M_{\odot}$ 
(\citealt{Hop09,Bez09}). According to these studies, however, it is not clear 
whether stellar mass density profiles evolve beyond the inner regions.
Whatever the case these results can only imply a similar or larger $\sigma$
at fixed $\Mstars$  at a higher redshift contradicting the abundance matching
results based on the downsizing SMF.  

What do hydrodynamic simulations predict on the evolution of the relation
between $\Mstars$ and $\sigma$?
\citet{Hop09b} combine galaxy merging with hydrodynamic simulation to find a 
little evolution of $\sigma$ with $z$ at fixed $\Mstars$ for any 
$10^9\Msun \le  \Mstars \le 10^{12} \Msun$. In particular,
\citet{Hop09b} explicitly predict that the $\Mstars$-$\sigma$ relation
evolves little between $z=1$ and $0$. \citet{Hop10} take into account a number
of possible effects in their cosmological simulations and find slow evolutions
of $\sigma$ with $z$. 
These results are consistent with the evolution in $\sigma(\Mstars)$ with
$z$ based on the COSMOS SMF  but not with the downsizing SMF.

{\it VDF evolution: concord or conflict with observed galaxy evolutions?}
We have already compared the evolving VDF with the evolving HMF and 
the evolving SMF. We find that the evolutions of the HMF, the VDF and the SMF
appear concordant, but that the downsizing SMF is disfavoured
because its implied structural evolutions are unlikely. 
 It is clearly worthwhile to put the VDF evolution in a
broader context of recent cosmological observations on galaxy evolutions.

The lensing constrained VDF evolutions show that the number density of
massive early-type galaxies ($ \sigma \ga 220 \kms$) not only evolves 
significantly but also shows a differential evolution (see Fig.~\ref{VDFs}): 
the higher the velocity dispersion, the faster the number density evolution 
(the ``velocity-upsizing'' behaviour), probably meaning an ``mass-upsizing'' 
behaviour (i.e.\ the behaviour of more massive galaxies assembling later in 
cosmic time). 
\citet{MK10} has just recently compiled a large number ($\sim 60,000$) of 
massive galaxies ($\Mstars > 10^{11} \Msun$) over a large sky area 
($55.2$~deg$^2$) from the UKIRT Infrared Deep Sky Survey (UKIDSS) and
the SDSS II Supernova Survey. \citet{MK10} find a significantly greater 
number density evolution for $\Mstars > 10^{11.5} \Msun$ than 
$\Mstars < 10^{11.5} \Msun$ out to $z=1$ consistent with the hierarchical 
evolution. The parallel evolution of the VDF and the SMF would imply no 
evolution in the total mass profile of galaxies. Indeed, strong lens modelling 
(\citealt{Koo06,SWF07,WRK04,TK02}) and velocity dispersion measurements 
to a high redshift (Fig.~\ref{Vzevol}) support non-evolution in 
total mass profiles even if stellar mass profiles evolve.
 Galaxy merging is an independent route to probe the build-up of 
galaxies over cosmic time (e.g.\ \citealt{Whi07,Mas08,Wak08,deR09,Bun09}). 
The mere fact that thousands of merging events
have been observed is the evidence for some sort of hierarchical mass assembly 
going on. The issue is the merging rate and its dependence on galaxy mass.
Observed merging rates are at variance and cannot test the hierarchical 
assembly of massive galaxies robustly. It is, however, worth noting the more
recent result by \citet{Bun09} that merging rate is greater for massive 
galaxies with $\Mstars > 10^{11} M_{\odot}$ than less massive galaxies.  
The \citet{Bun09} result is in line with the hierarchical mass assembly 
and agrees qualitatively with the VDF evolution.

\begin{figure}
\begin{center}
\setlength{\unitlength}{1cm}
\begin{picture}(8,7)(0,0)
\put(-1.2,7.3){\includegraphics{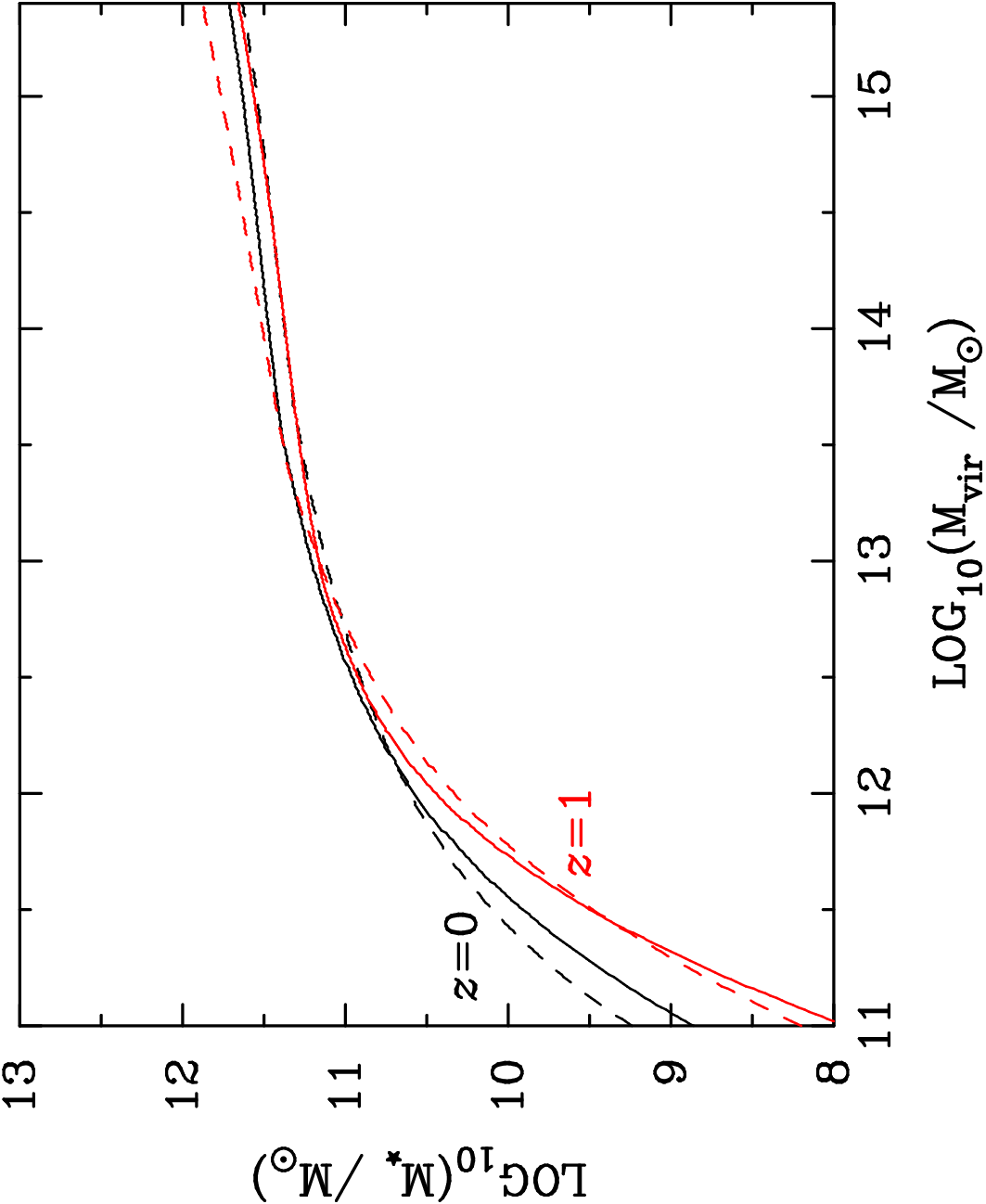}}
\end{picture}
\caption{
The abundance matching relations between $\Mvir$ (halo virial mass) and 
$\Mstars$ (central galaxy stellar mass) at $z=0$ (black curves) and $z=1$
(red curves) based on the COSMOS (solid curves) and the Spitzer (dashed curves)
 SMFs. These results are based on a constant intrinsic scatter of $0.16$ for 
$\log_{10}(\Mstars/\Msun)$ at fixed $\Mvir$. 
}
\label{MhMs}
\end{center}
\end{figure}

{\it Coevolution of $\Mvir$, $\sigma$ and $\Mstars$ and implications for the $\LCDM$ paradigm:} 
In the above discussions we have considered the connections of $\sigma$ with
 $\Mvir$ and $\Mstars$ separately. The results that $\sigma$ is coevolving in
parallel with both $\Mvir$ and $\Mstars$ necessarily imply 
a similar coevolution of $\Mvir$ and $\Mstars$. Fig.~\ref{MhMs} shows
the abundance matching $\Mvir$-$\Mstars$ relations at $z=0$ and $z=1$.
The results based on the COSMOS SMF give little evolution in 
the $\Mvir$-$\Mstars$ relation for $\Mvir \ga 10^{12} \Msun$ implying
a parallel coevolution of $\Mvir$ and $\Mstars$ with cosmic time. This is  
consistent with the little evolutions in the $\Mvir$-$\sigma$ and
the $\Mstars$-$\sigma$ relations based on the same SMF. Hence, 
we are left with the simple picture that $\Mvir$, $\Mstars$ and $\sigma$ 
are all coevolving so that a halo of given mass $\ga 10^{12} \Msun$ has on 
average the same stellar mass and the same stellar velocity dispersion for its 
central galaxy independent of redshift for $0 \la z \la 1$.   
Some indications of evolution in the $\Mvir$-$\Mstars$ relation (and
possibly in the $\Mstars$-$\sigma$ relation) for $\Mvir \la 10^{12} \Msun$ may
imply a differential evolution among $\Mvir$, $\Mstars$ and $\sigma$.
We cannot address this issue for low-mass haloes here because the strong 
lensing constrained VDF evolution from this work breaks down at low $\sigma$. 
The results based on the (downsizing) Spitzer SMF give a mild differential 
evolution in the $\Mvir$-$\Mstars$ relation at large 
masses\footnote{Given the observational uncertainty of the Spitzer SMF at $z=1$
 it is only marginally inconsistent with zero evolution (see \citealt{Beh10}).} 
as it is the case for $\Mstars$-$\sigma$ relation. 
However, in the above we have argued that a downsizing SMF is unlikely.

 How the parent dark halo is dynamically related to the residing galaxy as
a function of cosmic time is a fundamental question for galaxy formation and
evolution. In this work we have studied the connection of the halo mass 
($\Mvir$) with the stellar velocity dispersion ($\sigma$) and 
the stellar mass ($\Mstars$) of the central galaxy for $0\la z\la 1$.
According to our results the stellar dynamics in the galaxy (characterized
by $\sigma$) is closely linked to the parent halo mass $\Mvir$ independent of
redshift for $0\la z\la 1$. A link between 
the central particle velocity dispersion and $\Mvir$ is expected in the 
$\LCDM$ paradigm because it predicts on average a universal density profile
independent of $\Mvir$ (e.g.\ \citealt{Nav04,Nav10} and references theirin) 
along with well-defined scaling relations of the structural parameters with 
$\Mvir$ (e.g.\ \citealt{Bul01,Mac07,Kly10}). According to the coevolution,
 dissipational baryonic physics involving star formation that may have modified
the central potential of the halo has not destroyed but appears to have
refined the link.

It is then natural to suggest that dissipational baryonic physics results in
on average a rescaled universal (or universal class) density profile of the 
stellar plus dark mass distribution, or perhaps more realistically
a universal (class) density profile of dark matter combined with a 
well-correlated class of stellar mass distribution. What would then the 
baryon-modified universal (class) density profile look like? The observation 
that the inner density profile of the galaxy plus halo system is on average 
close to isothermal (see \citealt{Cha10} for a review and references)
combined with the expectation that baryonic effects are not likely to be 
important well outside the scale radii (e.g.\ \citealt{Gne04,Aba09,Tis10}) 
leads us to suggest a modified or generalized NFW (GNFW) profile in which
the inner total density profile is close to isothermal (with a possible
systematic variation with $\Mvir$) while the NFW profile
is kept at large radii well outside the stellar mass distributions.
The GNFW profile is then (in an average sense) preserved in the successive
merging of GNFWs.
Furthermore, the evolution of the concentration of such an GNFW profile
with $z$ at fixed $\Mvir$ conspires with the evolution of the virial radius
$\rvir$ with $z$ to lead to a non-varying $\sigma$ with $z$. It is not well 
understood at present whether this is just a coincidence or a revelation
of a fundamental mechanism in galaxy formation and evolution. It is also 
not clear whether the non-evolving $\Mvir-\sigma$ relation extends to
a higher redshift, i.e., since when the HMF and the VDF have been coevolving
in parallel in cosmic history.

The observation that the amount of star formation (i.e.\ stellar mass $\Mstars$)
is correlated with $\Mvir$ (e.g.\ \citealt{CW09,Mos09,Guo09,Beh10}) is 
consistent with the above picture. Namely, a larger halo undergoes 
a larger amount of star formation
needed to modify the greater potential well. If $\Mstars$ were perfectly
correlated with $\sigma$ at fixed $\Mvir$, the correlation between $\Mvir$
and $\sigma$ would be just a by-product of the $\Mvir$-$\Mstars$ correlation.
However, although there is some good correlation between $\Mstars$ and $\sigma$
for all galaxies (i.e.\ regardless of their haloes), the correlation
between $\Mstars$ and $\sigma$ at fixed $\Mvir$ is weaker (in preparation).
Hence, under the above picture the $\Mvir-\sigma$ correlation is originated
from the $\LCDM$ haloes and the amount of star formation set by $\Mvir$ 
rescales the correlation. We then expect some correlation between $\Mstars$ and 
$\sigma$ at fixed $\Mvir$ because the boost of $\sigma$ depends on the degree
of the baryonic effects on the halo characterized by $\Mstars$ (in preparation).

Let us compare the coevolution and the above picture motivated by it with 
some pictures (or interpretations) and numerical simulation results of the 
$\LCDM$ paradigm that have been discussed in the literature. 
First, the continual growth of the central stellar 
velocity dispersion and the stellar mass accompanying the growth of the halo 
over cosmic time would be inconsistent with a strictly ``stable core concept'' 
for massive galaxies (e.g.\ \citealt{LP03,Gao04}) even after $z=1$. 
However, the growth slopes for $\sigma$ and $\Mstars$ are shallower 
for more massive haloes according to the abundance matching results 
(see Fig.~\ref{MvirV} and Fig.~\ref{MhMs}). Hence
a weakly evolving core of massive haloes would be consistent with our results.
Second, the picture shares the concept of ``universal density profile'' 
with the attractor hypothesis (e.g.\ \citealt{LP03,Gao04}). However,
there is a clear distinction between the two. The attractor hypothesis assumes
that the universal NFW profile is preserved or restored in hierarchical 
merging of haloes (hosting galaxies) while the present picture assumes that the 
baryon-modified universal total density profile (i.e.\ the GNFW profile) is 
preserved once it is created. The latter property is supported by 
dissipationless merging simulations (e.g.\ \citealt{BM04,KZK06,NTB09}).

{\it Test of the $\LCDM$ paradigm?:} The basic tenet of the $\LCDM$ structure
formation theory is the hierarchical mass assembly. What the theory predicts
is the distribution of dark matter haloes. Connecting observed galaxies with 
theoretical dark haloes is a major goal of cosmological research. The 
difficulty of testing the $\LCDM$ paradigm arises from the complex physics of 
galaxy formation within the halo and the induced modification of the halo
structure. 
A necessary condition for a successful model is to reproduce the basic 
statistical properties of the observed local Universe, such as the luminosity,
stellar mass and velocity functions of galaxies and their correlations
(see \citealt{TG10}). However, a successful reproduction of the $z=0$ 
statistical properties of galaxies is not sufficient. A successful model
 must predict correctly the evolution of the galaxy properties. 
Semi-analytic models of galaxy formation have paid much attention on the 
galaxy luminosity and stellar mass functions. The current generation of these
models can reproduce the  $z=0$ functions reasonably well, but fail to match 
their observed evolutions (see, e.g., \citealt{Fon09,Str08,Cat08}).

The galaxy luminosity and stellar mass functions have much to do with the 
complex baryonic physics of star formations, AGN activities, feedbacks, etc.
Hence the evolutions of the galaxy luminosity and stellar mass functions can
only provide indirect tests of the underlying $\LCDM$ paradigm.
The stellar velocity or velocity dispersion of the galaxy residing in the
centre of a halo probes the gravitational potential of the baryon plus dark
matter system. Since the dark halo is expected to be modified in the course
of the dissipational galaxy formation process 
(e.g.\ \citealt{Blu86,Gne04,Rud08,Aba09,Tis10}) and the central 
potential is likely to be dominated by the baryonic matter, the velocity 
(dispersion) function evolution itself is not a direct probe of the $\LCDM$ 
paradigm either. However, the velocity (dispersion) function is separated from
much of the baryonic physics but has only to do with its dynamical effect.
Hence, once the dynamical effect of galaxy formation is well accounted for,
the evolution of the velocity (dispersion) function offers an useful
complementary probe of the structure formation theory. While it is challenging
to measure reliably the evolution of the velocity (dispersion) function 
through conventional galaxy surveys, strong lensing statistics in a well-defined
survey provides an excellent opportunity to constrain the evolution of 
the velocity (dispersion) function through the image splitting distributions 
(see \citealt{Cha10}). Current strong lensing statistics is 
limited by the small sample size. However, future cosmological surveys 
including (but not limited to) the Dark Energy Survey, the Large Synoptic
Survey Telescope and the Square Kilometre Array will increase dramatically 
the number of strong lenses (see \citealt{OM10}) allowing to put tight 
constraints on the evolution of velocity (dispersion) functions.

\section{Conclusions}

Through an abundance matching analysis of the lensing constrained VDF evolution
along with the theoretical HMF and the observed SMF from galaxy surveys, we find
 the following.

\begin{enumerate}

\item The dark halo virial mass-central stellar velocity dispersion 
($\Mvir$-$\sigma$) relation at $z=0$ is in excellent agreement with
 the observed properties of low-redshift  individual haloes.

\item The stellar mass-central stellar velocity dispersion 
($\Mstars$-$\sigma$) relation at $z=0$ is consistent with the local
scaling relations of galaxies in the literature.

\item The $\Mvir$-$\sigma$ relation  does not evolve 
between $z=1$ and $0$ independent of current observation and simulation data.

\item
The $\Mstars$-$\sigma$ relation does not evolve between $z=1$ and 
$0$ for the COSMOS SMF. This is well in line with the observed non-evolution of 
$\sigma$ with $z$ at $\Mstars = 10^{11} \Msun$. This is also consistent with
 the predicted little or mild evolution of $\sigma$ with $z$ 
insensitive to $\Mstars$ from cosmological simulations.
However, the Spitzer SMF (a typical downsizing SMF) requires the 
$\Mstars$-$\sigma$ relation to evolve in a differential way that is not 
supported by independent observational results on the structural evolutions
of galaxies in the literature.

\item The non-evolution in the $\Mvir$-$\sigma$ and the $\Mstars$-$\sigma$
relations imply a parallel coevolution of $\Mvir$, $\Mstars$ and $\sigma$
between $z=1$ and $0$.
This is corroborated by the little evolution in the abundance matching 
$\Mvir$-$\Mstars$ relation between $z=1$ and $0$ for $\Mvir \ga 10^{12}\Msun$.

\item The parallel coevolution of $\Mvir$, $\sigma$ and $\Mstars$ with $z$ 
may imply a universality and regularity in galaxy formation and evolution
despite complex baryonic physics processes. 

\end{enumerate}

\vspace{1cm}
The author would like to thank Mariangela Bernardi, Nacho Trujillo and 
Michele Cappellari for useful communications and Andrey Kravtsov, Robert 
Feldmann, Joshua Frieman, Nick Gnedin and Steve Kent for helpful discussions 
and conversations. The author also gratefully acknowledges the referee comments
 that were helpful in clarifying and improving the manuscript significantly. 

\bibliographystyle{mn2e}

\appendix

\section[]{Intrinsic scatters and bias corrections for the abundance matching relations}

This work is concerned with the abundance matching (AM) relations between 
$\sigma$ and $\Mvir$ and between $\sigma$ and $\Mstars$. The AM is
intended to recover the true median relation. If there were not any intrinsic 
scatter in the relation between two parameters, the AM 
by equation~(\ref{AM}) of two statistical functions would recover the true 
relation exactly. In reality the distribution of two observables in a plane has 
intrinsic scatters around the median relation. When there are such intrinsic 
scatters, the AM by equation~(\ref{AM}) will give a biased median relation that
is different from the true median relation. Here we estimate the bias and 
correct the AM relation by equation~(\ref{AM}) to obtain the corrected relation.
The bias-corrected AM relation is then checked for self-consistency through a 
Monte-Carlo simulation. In other words, we estimate the bias so that the 
corrected relation reproduces from one statistical function to the other  
through a Monte-Carlo simulation based on the intrinsic scatter.

For our purpose a knowledge of the intrinsic scatter is required.
There has been no measurement or simulation for the intrinsic scatter in the 
$\Mvir$-$\sigma$ relation. On the other hand, there have been measurements for 
the relation between $\Mstars$ and $\sigma$ (e.g.\ \citealt{Des07,HB09,Sha10}). 
Hence we study first the $\Mstars$-$\sigma$ relation using a model intrinsic
scatter motivated from observed intrinsic scatters. We then give some
 results on the $\Mvir$-$\sigma$ relation that come from a procedure to 
simultaneously derive an intrinsic scatter and the bias-corrected 
$\Mvir$-$\sigma$ relation by considering a bivariate distribution of $\Mstars$
and $\sigma$ as a function of $\Mvir$ (in preparation).
 Studies on the effects of the intrinsic scatters
in the $\Mvir$-$\Mstars$ and the $\Mvir$-$L$ relations can be found 
respectively in \citet{Beh10} and \citet{Tas04}.

\begin{figure}
\begin{center}
\setlength{\unitlength}{1cm}
\begin{picture}(8,6)(0,0)
\put(-1.,6.8){\includegraphics{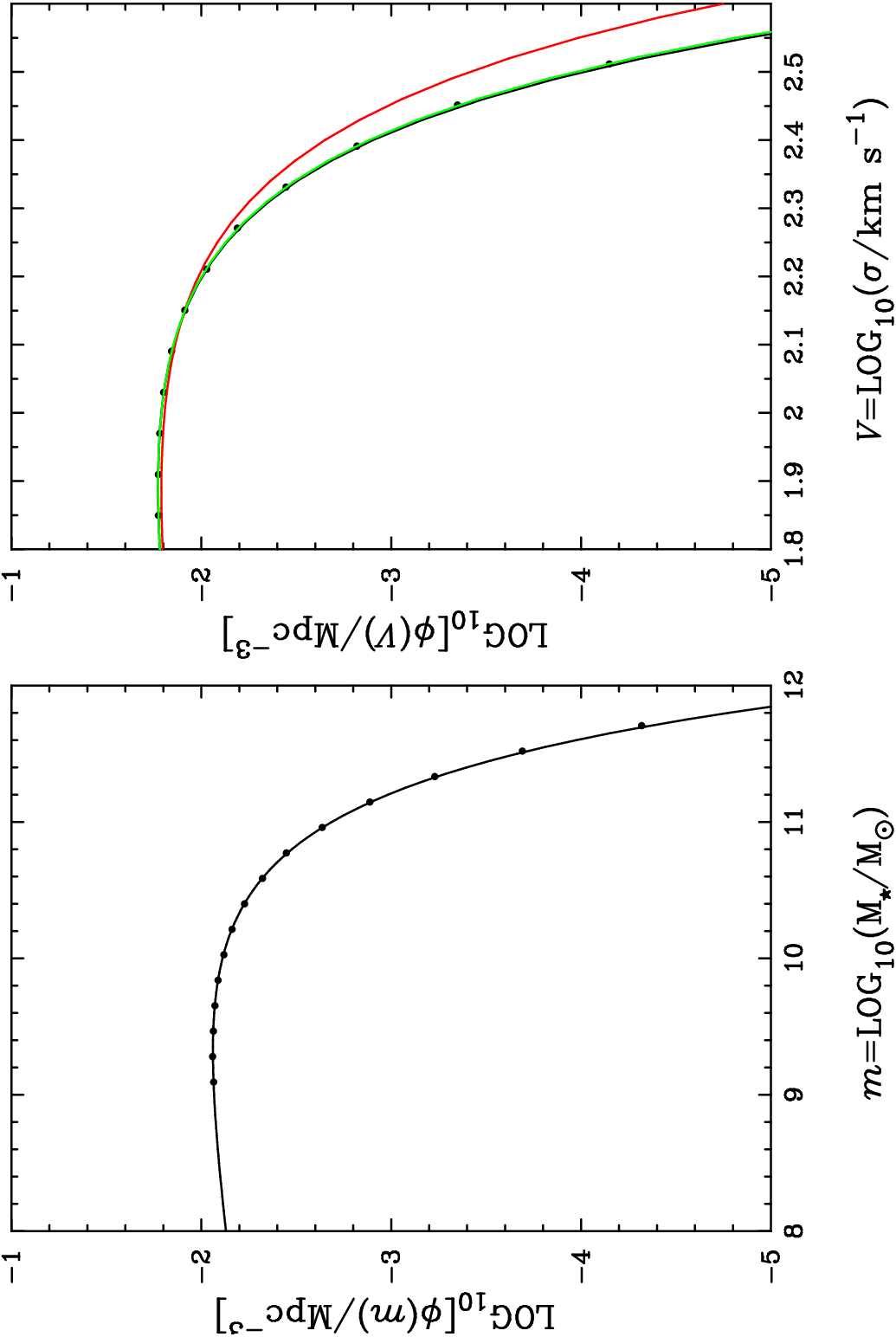}}
\end{picture}
\caption{
Left panel: The SMF for all galaxies measured by \citet{Ber10} based on SDSS 
DR6 and data points realised by a Monte-Carlo simulation. 
Right panel: The black curve is the VDF 
measured by \citet{Ber10} for $\sigma > 125\kms$ based on SDSS DR6.
 The data points and green curve are the results of a Monte-Carlo simulation 
based on the bias-corrected AM relation and the intrinsic scatter shown in 
Fig.~\ref{A2}. The red curve is the simulation result based on the
biased AM relation shown in Fig.~\ref{A2}.
}
\label{A1}
\end{center}
\end{figure}

For the purpose of demonstration we use the SDSS SMF and VDF by \citet{Ber10}
that are displayed in Fig.~\ref{A1}. For the intrinsic scatter of 
$V [\equiv \log_{10}(\sigma/ \kms)]$ as a function of 
$m [\equiv \log_{10}(\Mstars/ \Msun)]$, we adopt a linear model
given by $0.115 - 0.039 \times (m-10)$ that is derived from the \citet{Des07}
measurements of early-type galaxies for $m \ga 10.4$ and are consistent with
the measurements by \citet{HB09} and \citet{Sha10}. For all galaxies including
late-type galaxies, the intrinsic scatter will be more complicated than this.
In this work we do not attempt to consider an intrinsic scatter distribution for
all galaxies for the following two reasons. First, there have not been any
published measurement results of the intrinsic scatter for all galaxies. 
Second, the intrinsic scatter for early-type galaxies will match that for all 
galaxies at large stellar masses where the bias in AM is most significant. 
In other words, we can reliably estimate the greatest bias in 
AM using only the intrinsic scatter for early-type galaxies.

\begin{figure}
\begin{center}
\setlength{\unitlength}{1cm}
\begin{picture}(9,11)(0,0)
\put(-0.7,-0.4){\includegraphics{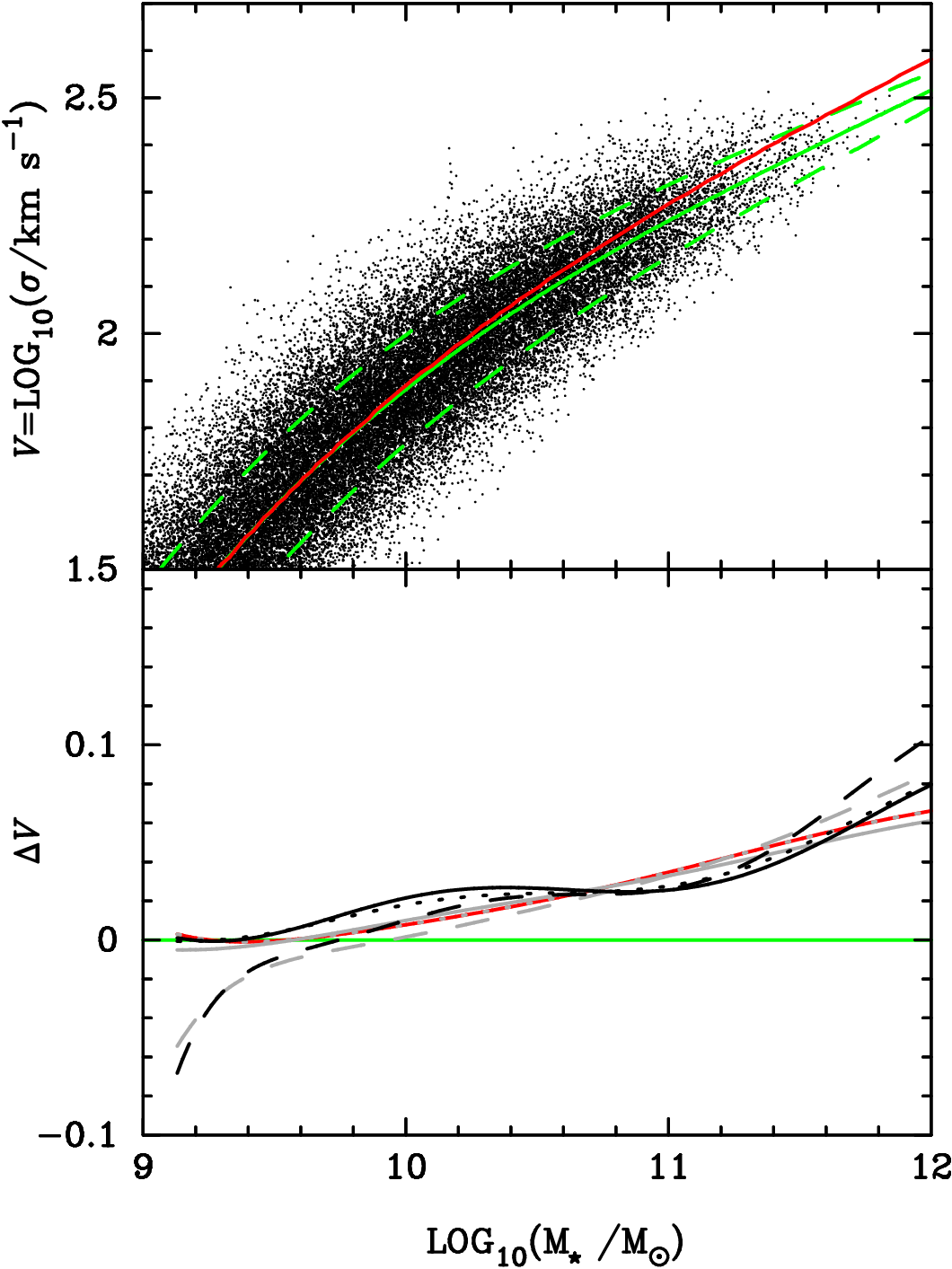}}
\end{picture}
\caption{Upper panel: 
Red curve is the AM relation given by equation~\ref{AM} for the SMF and the
VDF shown in Fig.~\ref{A1} while green curve is an adjusted relation. Green
dashed curves represent $1\sigma$ dispersion assuming a gaussian distribution
of $V[\equiv \log_{10}(\sigma/\kms)]$. Data points have been realized from
the SMF using the adjusted relation and the adopted scatter. A VDF derived 
from these simulated data points matches nearly perfectly the input VDF.
The adjusted relation  is referred to as the bias-corrected relation in
the texts. Lower panel: The difference between the initial AM relation 
(equation~\ref{AM}) and the adjusted relation, referred to as the AM bias.
Red curve is that for the upper panel. Other curves are for the following
input SMFs and VDFs: (1) black solid - Chae VDF/COSMOS SMF; (2) black dashed -
Chae VDF/Spitzer SMF; (3) black dotted - Chae VDF/Bernardi SMF;
(4) gray solid - Bernardi VDF/COSMOS SMF; (5) gray dashed -
Bernardi VDF/Spitzer SMF; (6) gray dotted - Bernardi VDF/Bernardi SMF 
(identical to the red curve).
}
\label{A2}
\end{center}
\end{figure}

Fig.~\ref{A2} shows the biased (red curve) and the bias-corrected (green curve)
AM relations. The dashed curves around the bias-corrected AM relation represent
the adopted intrinsic scatter described above. Fig.~\ref{A1} shows the input
VDF and the Monte-Carlo simulated VDFs from the input SMF based on the biased 
and bias-corrected AM relations and the adopted intrinsic scatter. Notice that
for the bias-corrected AM relation the simulated VDF closely matches the
input VDF. Hence the required self-consistency is gained. If we use
different input SMFs and VDFs, we will of course get different AM biases. 
Several biases required for the $z=0$ SMFs and VDFs used in this work 
can be found in the bottom panel of Fig.~\ref{A2}. To estimate the biases
required for the $z=1$ functions we use the predicted VDFs at $z=1$ 
based on the best-fit evolution parameters from strong lensing statistics 
along with the observed $z=1$ SMFs. The magnitude of the biases for $z=1$ is 
about the same as that for $z=0$ and we do not display the $z=1$ biases.

The same procedure would be followed for the $\Mvir$-$\sigma$ relation if
there were an observational intrinsic scatter as for the $\Mstars$-$\sigma$
relation above. Without a knowledge of the intrinsic scatter for the 
$\Mvir$-$\sigma$ relation we devise a procedure that allows us to derive
simultaneously the intrinsic scatter and the median relation for $\Mvir$
and $\sigma$ (in preparation). A full description of the procedure is
beyond the scope of this paper and the reader is referred to a following
paper in preparation. Here we only give a brief description of
the procedure and quote a simple result. The idea is to use a bivariate
distribution of $\Mstars$ and $\sigma$ at fixed $\Mvir$ noticing that 
$\Mstars$ and $\sigma$ are expected to be correlated. Then, we determine
simultaneously the scatter of $\sigma$ and the correlation coefficient between
 $\Mstars$ and $\sigma$ given the observed scatter of $\Mstars$ at fixed 
$\Mvir$ so that the resulting $\Mstars$-$\sigma$ relation is consistent with 
the given relation based on observations. The results depend on the input 
scatter of $\Mstars$ and the input $\Mstars$-$\sigma$ relation. 
Fig.~\ref{MvirV} shows a simple result 
based on a standard deviation of 0.16 for $\log_{10}(\Mstars)$ at fixed $\Mvir$
and the above described $\Mstars$-$\sigma$ relation based on the \citet{Ber10}
observational results.

\section[]{The $\vvir$-$\vopt$ relation at $z=0$}

\citet{Dut10} constrain the relation between $\vvir$ and $\vopt$ through
combining observationally derived $\Mstars$-$\Mvir$  and
$\Mstars$-$\vopt$ relations ($\vvir$ and $\vopt$ are the circular
rotation velocities at the virial and the optical radii respectively).  
The abundance matching $\Mvir$-$\sigma$ relation from this work 
 may be transformed into a $\vvir$-$\vopt$ relation using an
empirical relation between $\vopt$ and $\sigma$.  The virial velocity $\vvir$ 
is defined by $\sqrt{G \Mvir /\rvir}$ where the virial radius at $z=0$ is 
given by (\citealt{BN98})
\beq
\rvir \approx 209 h^{-1} \left(\frac{\Mvir}{10^{12} h^{-1} \Msun}\right)^{1/3}  
{\mbox{kpc}} 
\eeq
for the adopted cosmology and we take $h=0.7$.

For early-type (elliptical and  lenticular) galaxies, the direct estimate of
$\vopt/\sigma$ ranges from $\approx \sqrt{2}$  to 
$\approx 1.7$ (e.g.\ \citealt{Cou07,Ho07,Piz05,Fer02}). 
For late-type galaxies the \citet{Cha10} late-type VDF has actually been 
transformed from a circular velocity function (VF) assuming 
$\vopt/\sigma = \sqrt{2}$. Hence we can transform the \citet{Cha10} late-type 
VDF back to the original VF using the same factor. The \citet{Ber10} VDF is a 
directly measured function based on SDSS spectroscopy. It is important to note
that the SDSS measured velocity dispersions for late-type galaxies with small 
bulges (or without bulges)  come mostly from rotational motions (M. Bernardi, 
private communications), meaning that these small-bulge (bulgeless) systems are
 not missed in the \citet{Ber10} VDF.    Hence we need an independent
knowledge of $\vopt/\sigma $ for SDSS galaxies to transform the \citet{Ber10} 
VDF to a VF. Without a measured value of $\vopt/\sigma$ for SDSS galaxies we 
must resort to other measurements. From the literature  we find 
$\vopt/\sigma \approx 1.4 - 2$ for late-type galaxies depending on the 
bulge-to-disk ratio  (e.g.\ \citealt{Cou07,Ho07,Piz05,Fer02}). 

\begin{figure}
\begin{center}
\setlength{\unitlength}{1cm}
\begin{picture}(9,8)(0,0)
\put(-0.7,-2.2){\includegraphics{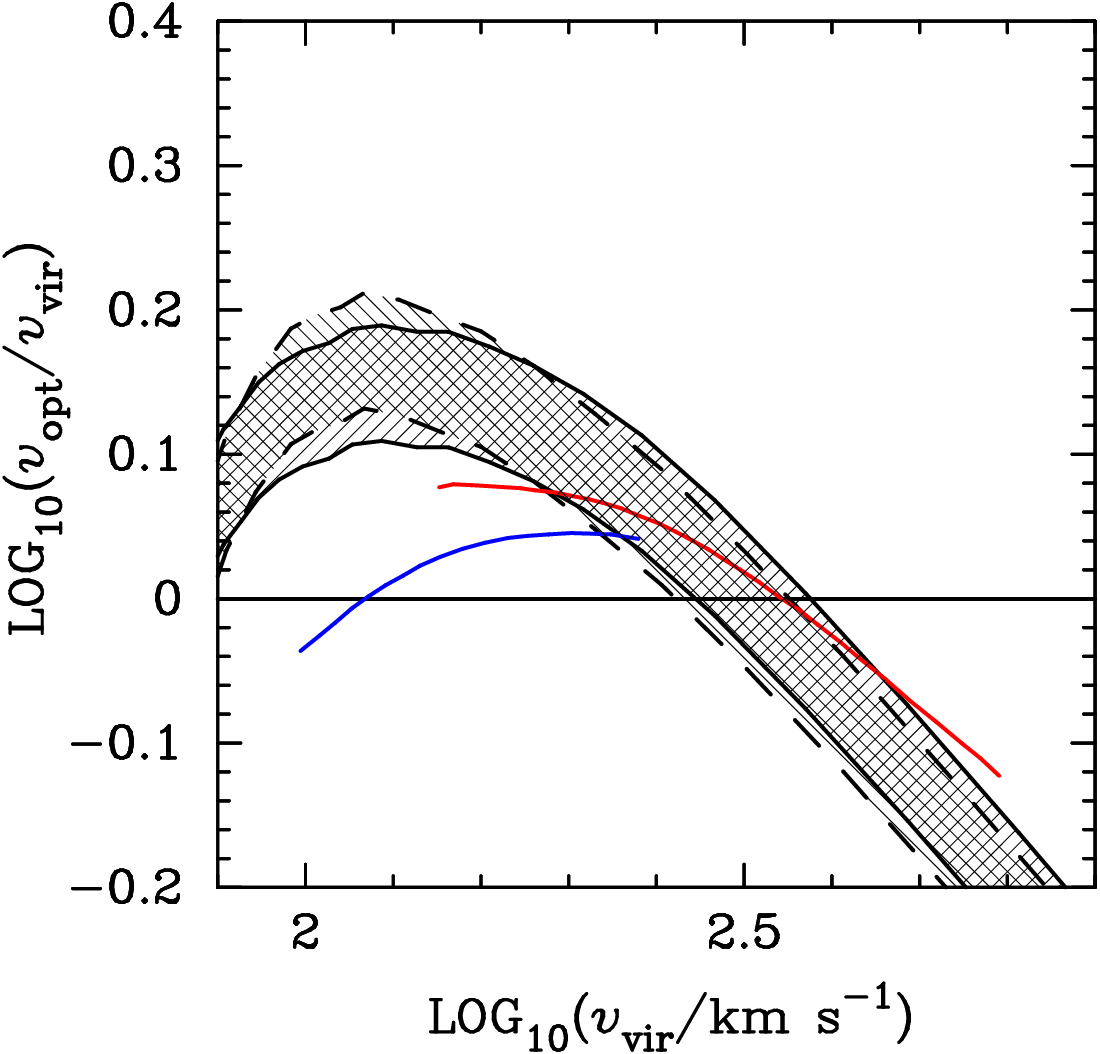}}
\end{picture}
\caption{
The relation between the halo circular  velocity at the virial radius ($\vvir$) 
and the stellar circular velocity in the optical region ($\vopt$) of the central
galaxy at $z=0$, inferred from the abundance matching relation between
$\Mvir$ (halo virial mass) and $\sigma$ (central stellar velocity dispersion)
from the VDF and the $\LCDM$ halo mass function (see the texts in \S 3.2).
The solid and dashed curves are respectively based on the VDFs by
 \citet{Cha10} and \citet{Ber10}. For each set of the results the lower and 
upper curves are respectively based on $\vopt=\sqrt{2}\sigma$ and 
$\vopt=1.7\sigma$. 
}
\label{MvirVoptz0}
\end{center}
\end{figure}

Based on these literature values of  $\vopt/\sigma $ for early- and 
late-type galaxies we adopt a range of  $\vopt/\sigma = \sqrt{2} - 1.7$ 
independent of galaxy type to estimate the circular velocity function of 
galaxies and then the $\vvir$-$\vopt$ relation through abundance matching.
Fig.~\ref{MvirVoptz0} shows the likely range of the median value for 
$\vopt/\vvir$ as a function of $\vvir$. 
Compared with Fig.~5 of \citet{Dut10} our results for 
massive galaxies ($\vvir \ga 10^{2.3} \approx 200 \kms$) overlap with the 
\citet{Dut10} results for early-type galaxies. 
For $\vvir \ga 10^{2.3} \kms$ our results give a scaling of 
$\vopt \propto \vvir^\gamma$ with $\gamma \approx 0.3-0.4$. 
This scaling is consistent with previous results for massive galaxies
based on strong lensing statistics and other methods (see \citealt{Cha06} and 
references theirin).

 However, our results for less massive galaxies ($\vvir \la 10^{2.3} \kms$)  
give systematically higher median values of $\vopt/\vvir$ compared with the 
\citet{Dut10} results for late-type galaxies although the estimated intrinsic
 scatters (not shown here) overlap. Our results are 
consistent with declining rotation curves near the virial radii for 
late-type galaxies while the \citet{Dut10} results imply flat 
rotation curves right up to the virial radii. The possible discrepancy may 
imply some unidentified systematic errors in one or both of the results. 
For our results the possible sources of systematic errors include the adopted 
VDFs and the adopted relation between $\sigma$ and $\vopt$.  
In order to be consistent with the \citet{Dut10} results 
the number densities of galaxies would have to be
significantly lower at relatively low velocity dispersions. We use two 
independently determined VDFs (i.e.\ the \citealt{Cha10} and \citealt{Ber10} 
VDFs) and they give similar results for $\vopt/\vvir$. In fact, the adopted
\citet{Ber10} VDF is the modified Schechter fit result for $\sigma > 125 \kms$.
This VDF gives an underestimate of galaxy number densities at low $\sigma$
according to \citet{Ber10} data. Hence if we were using the raw galaxy
number densities, the discrepancy with the \citet{Dut10} 
results would get worse  at low $\sigma$. 
The adopted range $\vopt/\sigma = \sqrt{2} - 1.7$ is suggested by a broad range
 of observations. A possible cause of error may be that bulgeless late-type 
galaxies have $\vopt/\sigma > 1.7$ so that we need to adjust the overall value 
higher. However, if we adopted $\vopt/\sigma > 1.7$, 
the discrepancy with the \citet{Dut10} results would get worse.

\end{document}